\documentclass{egpubl}
\usepackage{hpg25DL}

\WsPaper           %

\usepackage[T1]{fontenc}
\usepackage{dfadobe}

\usepackage{cite}  %
\BibtexOrBiblatex
\electronicVersion
\PrintedOrElectronic
\ifpdf \usepackage[pdftex]{graphicx} \pdfcompresslevel=9
\else \usepackage[dvips]{graphicx} \fi

\usepackage{egweblnk} 

\usepackage{booktabs}
\usepackage{makecell}
\usepackage{subcaption}
\usepackage{multirow}
\usepackage[usenames, dvipsnames]{color}
\usepackage{xcolor}
\usepackage{colortbl}
\usepackage{float}
\usepackage[export]{adjustbox}
\usepackage{listings}
\usepackage{algorithm2e}[noline]
\usepackage{amsmath}

\usepackage[commandnameprefix=always,defaultcolor=magenta,final]{changes}

\usepackage{xr}

\newcommand\SupplementaryMaterials{%
  \xdef\presupfigures{\arabic{figure}}%
  \xdef\presupsections{\arabic{section}}%
  \xdef\presupeqs{\arabic{equation}}%
  \xdef\presuptables{\arabic{table}}%
  \renewcommand\thefigure{S\fpeval{\arabic{figure}-\presupfigures}}
  \renewcommand\thesection{S\fpeval{\arabic{section}-\presupsections}}
  \renewcommand\theequation{S\fpeval{\arabic{equation}-\presupeqs}}
  \renewcommand\thetable{S\fpeval{\arabic{table}-\presuptables}}
}

\usepackage{xspace}
\newcommand{\FLIP}{\protect\reflectbox{F}LIP\xspace}

\newcommand{\WL}{List Merge\xspace}	%
\newcommand{\WB}{Box Sampling\xspace}	%
\newcommand{\WM}{Mask Sampling\xspace}	%
\newcommand{\WC}{Wave Communication STF\xspace}	%
\newcommand{\STF}{One-tap STF\xspace} %
\newcommand{\WLc}{List Merge\xspace}	%
\newcommand{\WBc}{Box Sampling\xspace}	%
\newcommand{\WMc}{Mask Sampling\xspace}	%

\title[Collaborative Texture Filtering]%
{Collaborative Texture Filtering}

\author[T.\ Akenine{-}M{\"o}ller, P.\ Ebelin, M.\ Pharr, B.\ Wronski]
{\parbox{\textwidth}{\centering 
		T.\ Akenine-M\"oller\orcid{0000-0001-6226-3170},
		P.\ Ebelin\orcid{0000-0003-3497-2943},
		M.\ Pharr\orcid{0000-0002-0566-8291},
		and	B.\ Wronski\orcid{0009-0005-0806-2307}
	}
	\\
	{\parbox{\textwidth}{\centering NVIDIA\\
		}
	}
}

\begin{document}

\teaser{
\vspace*{5mm}
	\setlength{\tabcolsep}{0.75pt}
	\renewcommand{\arraystretch}{0.5}	
	\newcommand{\mywidthbig}{30.8mm}
	\newcommand{\mywidth}{14.9mm}
	\newcommand{\outerLineWidth}{0.9pt}
	\newcommand{\innerLineWidth}{0.4pt}
	\newcommand{\grayLevel}{75}
	\vspace*{-5mm}
	\includegraphics[width=\mywidthbig]{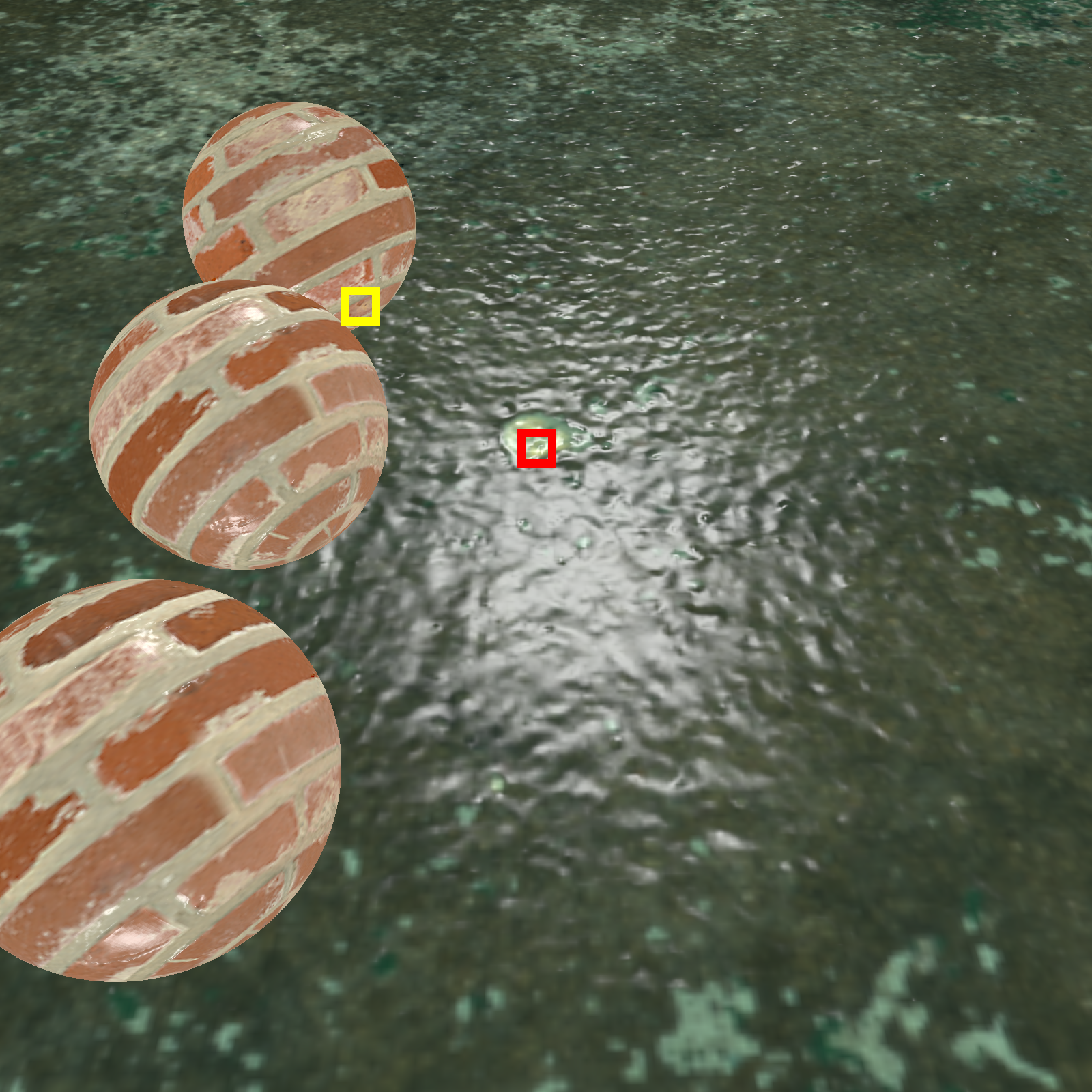}
	\begin{tabular}{ccc!{\color{gray!\grayLevel}\vrule width \outerLineWidth}cc!{\color{gray!\grayLevel}\vrule width \innerLineWidth}cc!{\color{gray!\grayLevel}\vrule width \innerLineWidth}cc!{\color{gray!\grayLevel}\vrule width \outerLineWidth}cc}
		\vspace{-34mm} &  \\
		&
		\multicolumn{2}{c!{\color{gray!\grayLevel}\vrule width \outerLineWidth}}{Ground truth} &
		\multicolumn{2}{c!{\color{gray!\grayLevel}\vrule width \innerLineWidth}}{\STF} &
		\multicolumn{2}{c!{\color{gray!\grayLevel}\vrule width \innerLineWidth}}{Wave Comm.\ STF} &
		\multicolumn{2}{c!{\color{gray!\grayLevel}\vrule width \outerLineWidth}}{Ours} &
		\\
		& 		
		\multicolumn{2}{c!{\color{gray!\grayLevel}\vrule width \outerLineWidth}}{PSNR / \FLIP~/ CVVDP} &
		\multicolumn{2}{c!{\color{gray!\grayLevel}\vrule width \innerLineWidth}}{28.7 /  0.045 / 9.18} &
		\multicolumn{2}{c!{\color{gray!\grayLevel}\vrule width \innerLineWidth}}{37.0 / 0.033 / 9.55} &
		\multicolumn{2}{c!{\color{gray!\grayLevel}\vrule width \outerLineWidth}}{\textbf{85.4 / 0.0000033 / 10.0}} &
		\includegraphics[width=\mywidth]{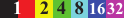}
		\\
		\rotatebox{90}{{\scriptsize not denoised}}  &
		\includegraphics[width=\mywidth,cfbox=red 1pt 0pt]{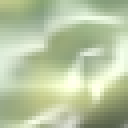} & 
		\includegraphics[width=\mywidth,cfbox=yellow 1pt 0pt]{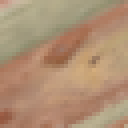} &
		\includegraphics[width=\mywidth]{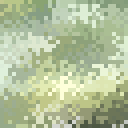} & 
		\includegraphics[width=\mywidth]{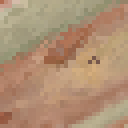} &
		\includegraphics[width=\mywidth]{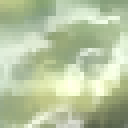} & 
		\includegraphics[width=\mywidth]{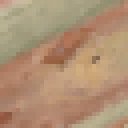} &
		\includegraphics[width=\mywidth]{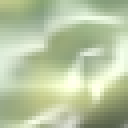} & 
		\includegraphics[width=\mywidth]{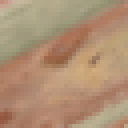} &
		\includegraphics[width=\mywidth]{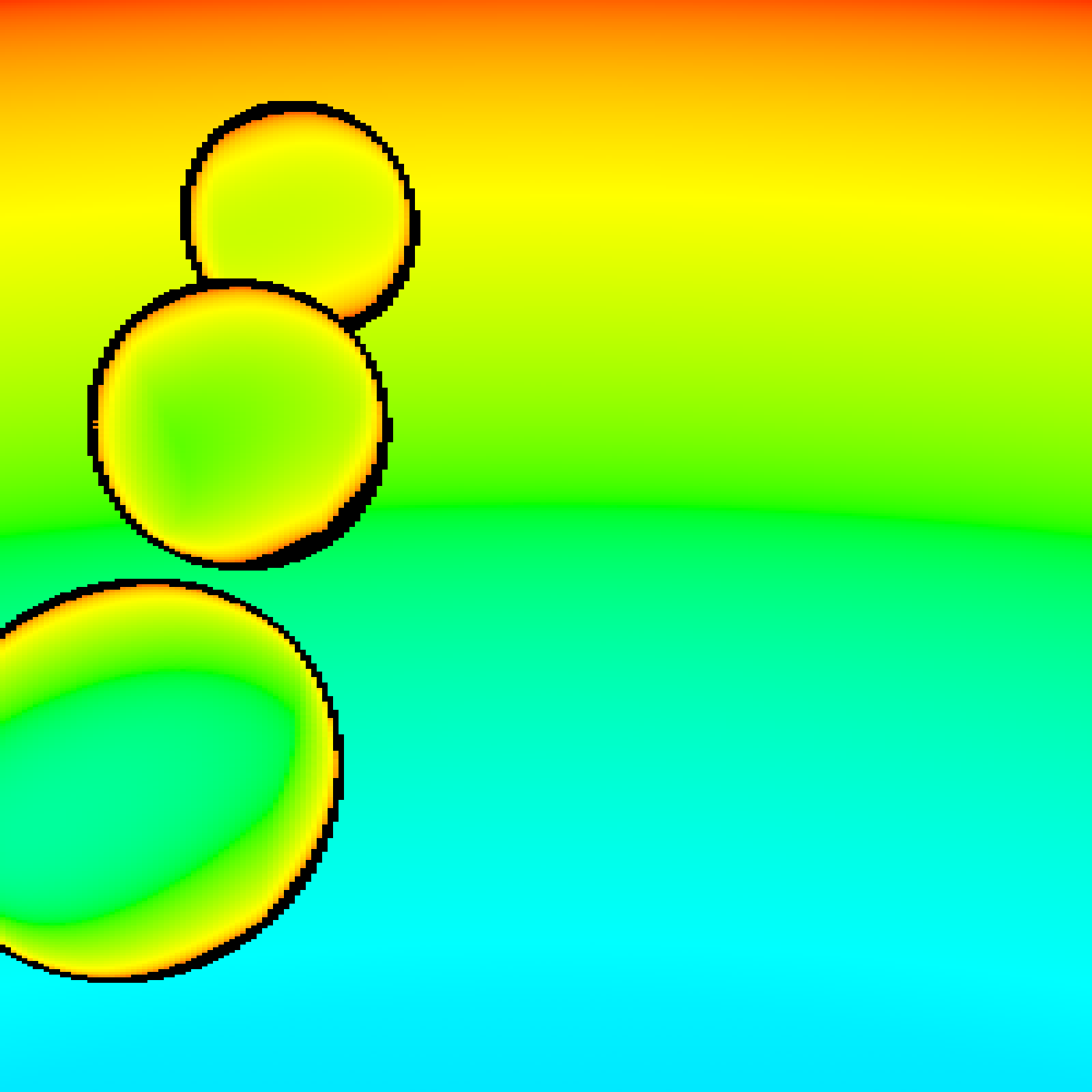}&
		\rotatebox{90}{{\scriptsize magnification}}
		\\
		\rotatebox{90}{{\scriptsize denoised}}  &
		\includegraphics[width=\mywidth,cfbox=red 1pt 0pt]{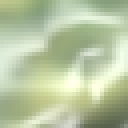} & 
		\includegraphics[width=\mywidth,cfbox=yellow 1pt 0pt]{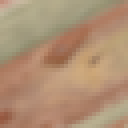} &
		\includegraphics[width=\mywidth]{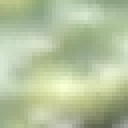} & 
		\includegraphics[width=\mywidth]{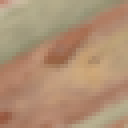} &
		\includegraphics[width=\mywidth]{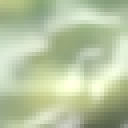} & 
		\includegraphics[width=\mywidth]{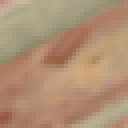} &
		\includegraphics[width=\mywidth]{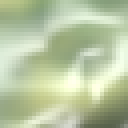} & 
		\includegraphics[width=\mywidth]{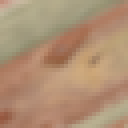} &
		\includegraphics[width=\mywidth]{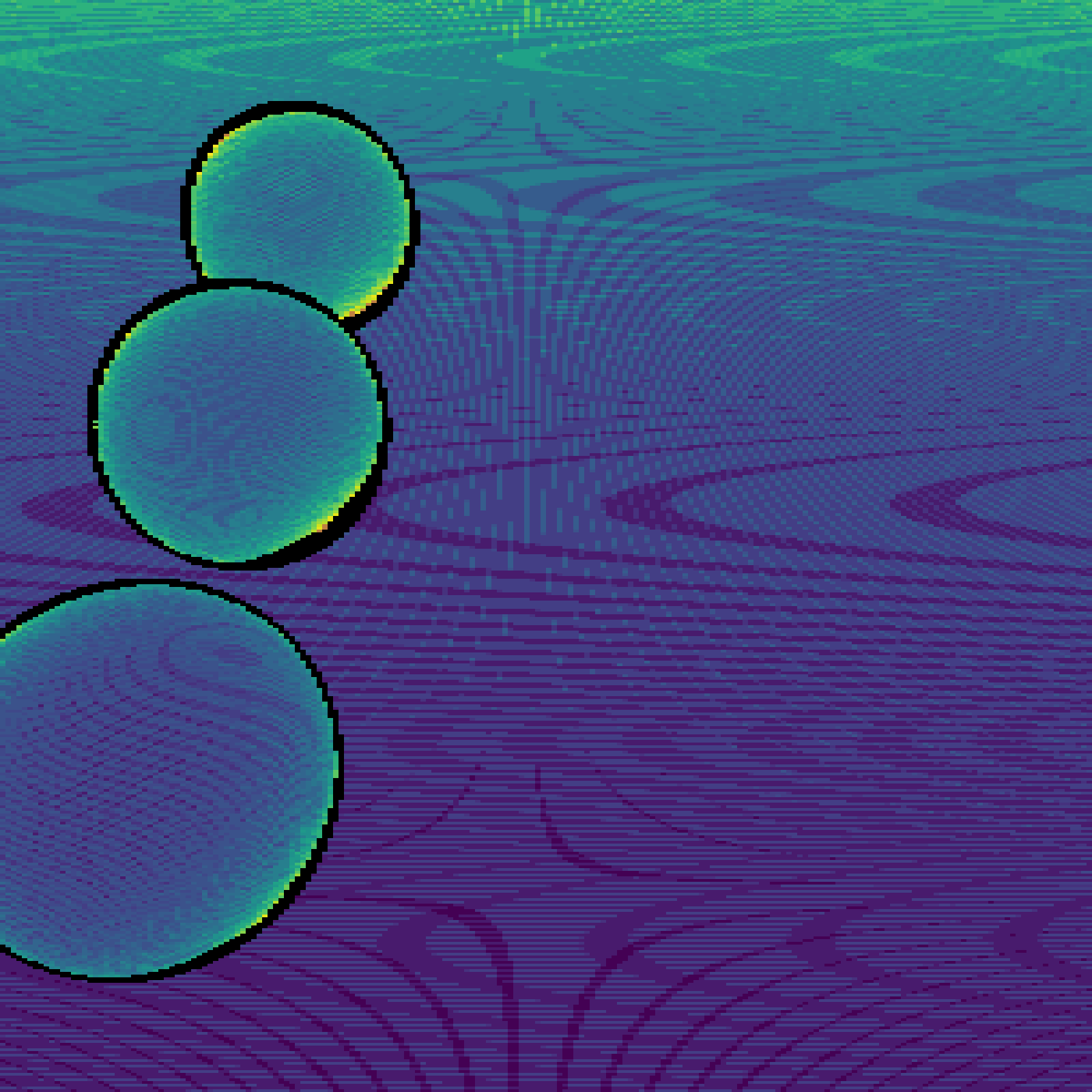}	&
		\rotatebox{90}{{\scriptsize texel accesses}}
		\\
		\vspace*{-1.8mm}
		\\
		&
		\multicolumn{2}{c!{\color{gray!\grayLevel}\vrule width \outerLineWidth}}{PSNR / \FLIP~/ CVVDP} &
		\multicolumn{2}{c!{\color{gray!\grayLevel}\vrule width \innerLineWidth}}{37.4 / 0.033 / 9.36} &
		\multicolumn{2}{c!{\color{gray!\grayLevel}\vrule width \innerLineWidth}}{45.8 / 0.019 / 9.87} &
		\multicolumn{2}{c!{\color{gray!\grayLevel}\vrule width \outerLineWidth}}{\textbf{71.2 / 0.0013 / 10.0}} &
		\includegraphics[width=\mywidth]{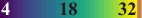}
	\end{tabular}
	\caption{
		During texture magnification, where previously only
		stochastic texture filtering (STF) was a viable option due to the expense of \chreplaced{texel}{texture} evaluation,
		our algorithms often produce zero-error results at negligible cost.
        We compare original \STF~\cite{Pharr2024} and \WC~\cite{Wronski2025}
        to our approach, both with and without DLSS~\cite{NVIDIA2025:DLSS}.
		For this scene, one of our methods uses only \chhighlight{$0.34$}\chdeleted{$0.3$} \chreplaced{texel evaluations}{texture lookups}
		per pixel on average, compared to 4 for classic bilinear interpolation, and yet
		renders an image with extremely low error.
		On the right, \chadded{we show false color visualizations of the scene, where}
		\chhighlight{the color maps use black to indicate waves with at least one pixel
		that uses minification}.
		The top image shows magnification
		factors ranging from \chhighlight{$\sim 0.0$}\chdeleted{$0.007$} to \chhighlight{$9.1$}\chdeleted{$22$} (average: \chhighlight{4.3}\chdeleted{$9.4$})
		and the bottom image illustrates how many unique texels are needed per wave
		($8\times 4$ pixels) to achieve perfect bilinear filtering.
		The \chadded{bottom} color map uses brown to 
		indicate waves where this is not reached with $\leq 1$ \chreplaced{texel evaluation}{lookup} per pixel.
		For all the other waves, our method \chreplaced{produces}{fetches} all $2\times 2$ texels
		needed for perfect bilinear filtering.
	}
	\label{fig_teaser}
}

\maketitle
\begin{abstract}
	Recent advances in texture compression provide major improvements in compression ratios,
	but cannot use the GPU's texture units for decompression and filtering.
	This has led to the development of stochastic texture filtering (STF) techniques
	to avoid the high cost of multiple \chreplaced{texel evaluations}{texture lookups} with such formats.
	Unfortunately, those methods can \chreplaced{give}{result in} undesirable visual appearance changes under magnification
	and may contain
	visible noise and flicker despite the use of spatiotemporal denoisers.
	Recent work substantially improves the quality of magnification filtering with STF by sharing decoded texel values between nearby pixels~\cite{Wronski2025}.
	Using GPU wave communication intrinsics, this sharing can be performed inside actively executing shaders without memory traffic overhead.
	We take this idea further and present novel algorithms that use wave communication between lanes 
	to avoid repeated texel decompression prior to filtering.
	By distributing unique work across lanes, we can achieve \emph{zero-error filtering}
	using $\leq 1$ \chreplaced{texel evaluations}{texture lookups} per pixel given a sufficient\chadded{ly large} magnification factor.
	For the remaining \chreplaced{cases}{situations},
	we propose novel filtering fallback methods that also achieve higher quality than prior approaches.

		
\end{abstract}  

\section{Introduction}
\label{sec_intro}

Stochastic texture filtering (STF) has recently reemerged as a useful technique~\cite{Pharr2024} for filtering texture representations such as neural encodings~\cite{Vaidyanathan2023,Fujieda2024,Kim2024neuralvdb,Dupuy2025} %
that are expensive to evaluate.
Not only does it allow performing fewer texel evaluations than traditional deterministic texture filtering requires,
but STF also allows efficient implementation of \emph{filtering after shading}, where the filter is applied to the final shaded values rather than the texture input.
When shading functions have nonlinearities and textures are minified, filtering the final result can give higher-quality images.
However, when textures are magnified, filtering after shading can introduce aliasing~\cite{Pharr2024}; in that case filtering of the texel values is preferable.

With native GPU texture formats, accessing all the texels needed for the filter has
limited cost since 
the cache hierarchy %
is effective at greatly reducing the memory bandwidth used for redundant texture lookups at nearby pixels.
On the other hand, directly filtering custom texture formats in a shader is computationally expensive since each pixel must produce every texel it needs.
Caches give less benefit:
while they can still reduce the bandwidth, e.g., for reading network weights
used by multiple \chreplaced{texel}{texture} evaluations, they do not save the redundant computation.

With the \emph{\STF} algorithm, each pixel always samples a single texel and filtering after shading is always performed~\cite{Pharr2024}.
Recent work by Wronski et al.\ introduced a \emph{\WC} algorithm that uses GPU wave-intrinsic instructions to
communicate texel values among groups of pixels running in the same GPU wave~\cite{Wronski2025}.
This allows filtering before shading under magnification, giving results significantly closer to full texture filtering compared to \STF without increasing the number of texels evaluated.
However, with \WC, each pixel independently determines which texel to evaluate based on its own texture filter.
Although Wronski et al.\ introduced methods based on optimized blue noise samples and stochastic sharing masks to increase the probability of pixels finding novel texel samples at their neighbors, their method does not guarantee that nearby pixels will not redundantly evaluate the same texel. Under the GPU SIMT execution model, such redundant computation evaluation does not necessarily harm performance;
runtime is generally determined by the maximum computation done in any pixel in a wave, so having multiple lanes do the same computation has no additional runtime cost.
However, redundantly evaluating the same texel does not \chreplaced{give any}{add} new information necessary
for more accurate filtering; image quality could be further improved if one could guarantee
that different lanes produce unique texels.

In this work, we develop algorithms for \emph{collaborative texture filtering} (CTF) for
when textures are magnified,
where collections of pixels communicate to determine their collective texel requirements and then ensure that each pixel evaluates a unique texel.
Each pixel then gathers the texels it needs for its filter from its lane neighbors
through efficient wave-intrinsic instructions.
For moderate ($1.59\times$ or $2.35\times$, depending on which of our methods is used) or
greater magnification with 32-lane waves, our approach guarantees that each pixel has all the texels required for
perfect bilinear filtering. %
\chreplaced{For}{Under} lower magnification \chadded{factors} or with wider filters that require more texels, all texels may not be available.
In this case, we have each pixel prioritize texels based on their contribution to its filtered
result, similarly to \WC.
However, for such cases we propose a novel way of combining those texel values that
improves image quality compared to One-tap and \WC, with negligible additional computational cost.

\chreplaced{For}{Under} high magnification \chadded{factors}, only a few texels may be required by all pixels in a wave.
In this case, not only do we ensure that all pixels have all texels needed for filtering,
but there is an opportunity for improved performance if the computation for texel evaluation can be split across multiple lanes and performed in parallel.

Our primary contributions are:
\begin{itemize}
\item The introduction of three algorithms for collaborative texture filtering across lanes in a wave,
used to achieve perfect filtering under magnification: \emph{\WL}, \emph{\WB}, and \emph{\WM}, spanning different design points that trade off %
computation and the effectiveness of finding a minimal set of texels required.
\item New \emph{fallback} methods for when \chreplaced{our techniques above are}{CTF is} unable to provide all texels needed for perfect filtering,
which give higher quality output than \STF~\cite{Pharr2024} or \WC~\cite{Wronski2025}.
\item Evaluation of our algorithms with bilinear, bicubic B-spline, and Catmull--Rom filters,
comparing performance and image quality to \STF and \WC.
\item Single-tap filtering for filters with both negative and positive weights, such as the Catmull--Rom filter, without requiring more expensive techniques such as positivization~\cite{Pharr2024}.
\end{itemize}

\section{Previous Work}
\label{sec_prevwork}

Stochastic texture filtering has seen previous use dating to the 1990s; 
see Pharr et al.~\cite{Pharr2024}
for an extensive history.
When textures are magnified, the filtering after shading normally performed with STF can introduce aliasing; 
recent work by Wronski et al.~\cite{Wronski2025} addresses this issue by sharing texel samples across groups of nearby pixels.
When all texels required to filter a texture at a pixel are available, it is possible to perform traditional
filtering for magnified textures, eliminating this aliasing.
In general, even if only some of the additional texels are available, such texel sharing reduces error.

The sharing approach by Wronski et al.~is based on \chdeleted{efficient} \emph{wave intrinsics} that allow efficient sharing of values between shader instances at nearby pixels.
Introduced in DirectX HLSL Shader Model 6.0, wave intrinsics expose the notion of a \emph{wave} of individual \emph{lanes} that are executing shader instances as a group,
mirroring the underlying GPU execution model~\cite{Microsoft2021}. See the
paper by Wronski et al.\ for a more detailed introduction to the concept~\cite[Section 2.3]{Wronski2025}.
Because shader instances execute together, wave intrinsics can efficiently use vector registers for communication
instead of using local or off-chip memory.
(While lanes may be mapped to pixels, vertices, or other elements being processed by shaders, we will sometimes use ``pixels'' interchangeably with ``lanes.'')
Our work is inspired by that of Wronski et al. and our algorithms make similar use of wave intrinsics.

Communication of data between executing shader instances has been used for a variety of other high-performance graphics algorithms, including
in-place screen-space filtering~\cite{penner:2011:shader,mcguire:2012:scalable}.
Communication between shading passes is also frequently performed using off-chip memory;
examples include ReSTIR, which communicates samples between pixels, both spatially and temporally~\cite{bitterli:2020:restir}, and post-processing antialiasing techniques like TAA~\cite{Karis:2014:High,Yang:2020:Survey} and
DLSS~\cite{NVIDIA2025:DLSS} that temporally accumulate shaded pixel values.

The algorithms we introduce are based on pixels individually identifying texels required for filtering and calculating a shared set of texels that need to be produced for the wave.
When this set of texels does not exceed the number of pixels in the group, we distribute the texel decoding cost between different pixels to avoid redundant \chreplaced{texel evaluations}{texture lookups}.
Tiled and clustered lighting algorithms~\cite{olsson:2012:clustered} similarly communicate between local groups of pixels to determine their spatial and directional bounds.
Then, they use local groupshared memory to collaboratively cull the light list to determine a subset of lights that may contribute to the given pixel cluster's shading.
Also related are techniques like resolution-matched shadow maps where shadow map lookup requests are written to memory and
parallel compaction algorithms determine which shadow quadtree pages to render~\cite{lefohn:2007:rmsm}.
(Some virtual texturing algorithms follow similar approaches.)
\chadded{Kenzel et al. propose to avoid redundant work in compute-based rasterization pipelines by merging pixel coverage bit masks~\cite{kenzel2018high}.}
By using wave intrinsics rather than groupshared or off-chip memory to organize texel requests, our approach
achieves high performance and
does not require shaders to finish or synchronize execution before launching a separate set of shaders to process requests. 

A characteristic of our approach is that each pixel does not necessarily compute a texel value that it needs itself.
This mirrors a common GPU programming technique, where threads may not have a fixed association with data elements.
Classic examples are high-performance GPU scan and sorting algorithms~\cite{sengupta:2011:efficient,satish:2009:sort}.

\section{Collaborative Texture Filtering}
\label{sec_algorithm}

\begin{figure*}[t]
	\centering
	\includegraphics[width=\textwidth]{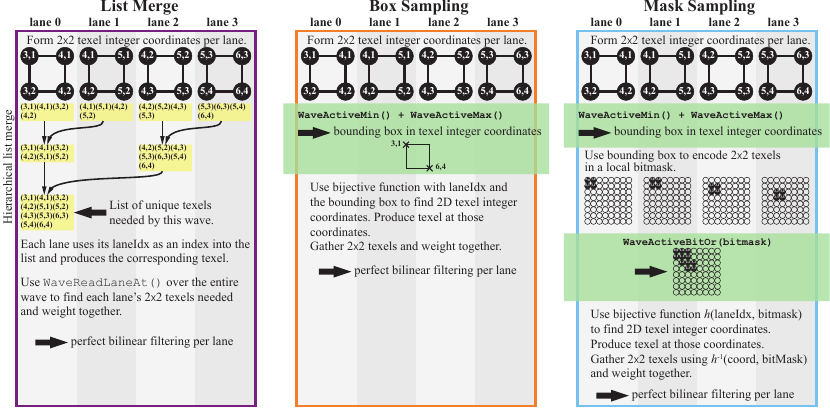}
	\caption{High-level overviews of our three algorithms for the bilinear filtering case,
		shown with only four lanes (in gray) and where algorithm flow is downward.
		See our pseudocode in Section~\ref{supp_pseudocode} for details of early outs to fallback methods.
		Shared memory is shown in yellow, while cross-wave operations are green. %
		\WL performs a hierarchical list merge on
		the $2\times 2$ texel integer coordinates per lane. This gives a
		list of unique texels needed by the wave to be able to perform
		bilinear filtering in each lane.
		\WB is our fastest technique and uses an efficient
		bijective function to find which texel is \chreplaced{produced}{needed} by which lane.
		\WM creates a local bitmask where each bit indicates
		a texel integer coordinate, and a set bit indicates that the texel is needed
		by the wave.
		\WM is exemplified using an $8\times 8$ bitmask.
	}
	\label{fig_algorithms}
\end{figure*}

In this section, we introduce a set of techniques that we call
\textit{collaborative texture filtering} (CTF). 
For simplicity, the majority of our description is related to bilinear
filtering, but as we will show later on, it also extends to 
arbitrary filters, e.g., bicubic B-spline and Catmull--Rom.
In addition, we assume that each wave has 32 lanes and is configured
as $8\times 4$ pixels. However, our methods generalize to any wave size
and can work even better at lower magnification factors with larger waves.
The key feature of all our algorithms is that they
aim to provide all the texels required for \textit{perfect}, i.e., with zero error,
filtering with at most one texel \chreplaced{evaluation}{lookup} per lane for all the lanes in a wave. 

All our new methods share the same general algorithm flow, which is summarized as:
\begin{enumerate}
	\item The lanes in the wave collaboratively collect information about which texels
		  are needed for perfect filtering.
	\item \chadded{Texel evaluation:} each lane \textit{produces} up to one texel based on the information from the previous step.
	\item Each lane \textit{gathers} the texels it needs from across the wave and filters them.
\end{enumerate}
In step~2, we use the term \textit{producing} a texel to include, e.g., a regular texel access using the GPU texture unit,
decompressing a texel stored in a custom format, procedural generation, or traversing a data structure to obtain the
texel~\cite{Kim2024neuralvdb}. Note also that we sometimes use
 ``produce a texel'' interchangeably with ``\chreplaced{texel evaluation}{texture lookup}.''

In some cases, e.g., when the number of unique texels from
step one is too large, a fallback method is used.
In general, one may fall back to any filtering method, such as \STF~\cite{Pharr2024}
or \WC~\cite{Wronski2025}, though
we present novel fallback methods in Section~\ref{sec_fallback_method} that
result in higher image quality. %

Next, we describe three implementations of the algorithm flow above:
\textit{\WL}, \textit{\WB}, and \textit{\WM}.
Each has different qualities, such as performance and success rate of perfect filtering.
A high-level description of them can be found in Figure~\ref{fig_algorithms}.%

\subsection{\WLc}
\label{sec_wl}
\WL creates a list of the unique texels required by the wave.
This list is constructed in shared memory
using a hierarchical list merge algorithm.
The texture coordinates in each lane determine which $N\times N$ texels
are needed for filtering its texture. For now, we will focus on the bilinear filter with $2\times 2$
texture filtering footprint.
In this case, the integer coordinates of these four texels are put in a list of length four in each lane.

As illustrated to the left in Figure~\ref{fig_algorithms},
a hierarchical list merge is then performed, using shared memory to store the lists during the merge.
This is similar to the hierarchical parallel sum described by Hoobler~\cite{Hoobler2011}\chreplaced{, though}{The difference is that} instead of summing two values, we merge two lists such
that the list never contains any duplicate texel coordinate pairs.

After the final list has been formed, we check
if the number of elements in the list is larger than
the number of available lanes. If so, perfect filtering \chadded{with
one texel evaluation per lane} is not possible 
and we revert
to a fallback method (Section~\ref{sec_fallback_method}).
Otherwise, each lane first
produces the texel whose index in the merged list is the same
as its lane index. After texels have been produced, each lane loops over
lanes in the wave to gather the texels that fall within
its filter footprint and then weights them using the original texture filter weights.
This method has substantially worse runtime cost than
our other methods, and therefore we omit further details.
We include it in the discussion because it provides an upper bound on the success rate
of when perfect filtering is possible and demonstrates the goal we aim to more efficiently approximate.

\subsection{\WBc}
\label{sec_wb}
On the opposite spectrum of complexity to \WL is a much simpler
alternative that we call \emph{\WB}, illustrated in the middle 
of Figure~\ref{fig_algorithms}.
In this method, each lane calculates an axis-aligned
bounding box (AABB) of the texels it needs for texture filtering\chdeleted{,
uniquely identified by its top left and bottom right corner}.
The size of this local \chdeleted{and initial} AABB depends on the %
texture filter---for example $2\times 2$ texels for bilinear
and $4\times 4$ for bicubic filtering.
We use those local bounds to compute a global AABB
for the wave using the \texttt{WaveActiveMin()} and \texttt{WaveActiveMax()}
intrinsics.

If the area $n$ of the global bounding box exceeds the number of active lanes in the wave,
we use one of the fallback methods (Section~\ref{sec_fallback_method}).
Otherwise, in cases where at least the first $n$ texels are active,\footnote{For cases when this
is not true, we refer to the description in our supplemental (Section~\ref{sec_remapping}), where
an additional mapping function is needed.}
we use a simple bijective function to map from the active lane index to \chdeleted{the $x$- and $y$-}coordinates
within the bounding box by using modulo and division by the bounding box width.
Each active lane with an index $<n$ uses this mapping to produce the corresponding texel from the texture.
Finally, each pixel uses \texttt{WaveReadLaneAt()} instructions with the source lane index computed through \chreplaced{inverse}{reverse} mapping from local \chdeleted{$xy$-}coordinates inside the AABB to
gather the texels needed for its filter,
and weights them to generate the pixel's perfectly-filtered value.

While \WB is substantially simpler and faster than \WL,
an AABB may include many unnecessary texels
under perspective projection or rotations around 45 degrees, resulting in more frequent need of the fallback method
and thus increased \chdeleted{image} error.
To address this limitation, we propose a slightly more complex hybrid method that increases the number of perfectly-filtered waves
without incurring the excessive cost of \WL.

\subsection{\WMc}
\label{sec_wm}
Similar to \WB, \WM starts by creating an AABB
over the texel integer coordinates.
If it is larger than $16\times16$ texels, then we use one of the fallback methods
(Section~\ref{sec_fallback_method}).
Otherwise, we use a $16\times 16$ bitmask to \chreplaced{encode which}{help with processing of
the needed} texels \chadded{are needed}. This bitmask is stored using an \texttt{uint64\_t4} variable,
i.e., $4\cdot 64 = 256 = 16\cdot 16$ bits\chdeleted{ for storing one bit per texel
inside the bounding box}.
Each lane \chdeleted{then }sets the four bits in the mask corresponding to the
$2\times 2$ texels it needs
for bilinear filtering.
We then use \texttt{WaveActiveBitOr()} on these masks to find a \textit{wave mask},
called $B$,
that contains a 1 for each texel the wave needs for perfect bilinear filtering.
If the number of bits set in the wave mask, $n$, is larger than 32,
we also call one of the fallback methods.

Otherwise, we know that there are $n$ unique texels in this wave and $n\leq 32$,
and for now we assume that the first $n$ lanes are active in the
wave. We refer to the description in our supplemental (Section~\ref{sec_remapping}) for cases when
this is not true.
If $i<n$, where $i$ is the current lane's index, then lane $i$ will produce a
texel.\chdeleted{and share it with the rest of the wave.}
One of the texels marked by a 1 in the wave mask needs to be selected.
Hence, a mapping $h$
is needed from lane index $i$ to \chadded{a requested} one-dimensional texel
coordinate\footnote{One can convert from two-dimensional coordinates,
$(t_x,t_y)$, to one-dimensional as $t = 16\cdot t_y + t_x$.} $t$
in the wave mask, $B$. 
We choose $h$ to map from $i$ to the $i$th bit that is set in $B$,
and call the function $t = h(i,B)$. The number $t$ is in $[0,255]$, and so
can be mapped to local two-dimensional coordinates in $[0,15]^2$ and 
\chdeleted{then} added to the upper left integer coordinates \chreplaced{of the AABB}{over the entire wave}.
These coordinates are then used to produce the desired texel.

\begin{figure}[t]
	\centering
	\includegraphics[width=\columnwidth]{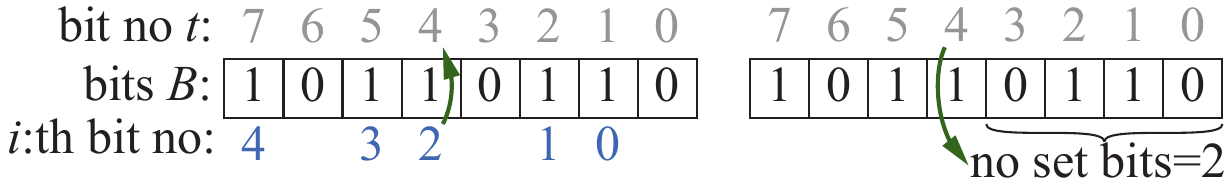}
	\caption{Illustration of our bijective function $t=h(i,B)$ used
		in \WM.
		Assume the lane index $i=2$, then $h(2, B)=4$ since the bit number of
		the 2nd set bit is 4, as illustrated by the left arrow. 
		Note that we start counting from 0,
		so the 0th set bit is at position 1 in the example above.
		To the right, we map from $t=4$ to lane index $i$ by
		counting the number of set bits to the right of
		$t=4$. The result is $2$, so $h^{-1}(4,B)=2$.
	}
	\label{fig_bijective_function_h}
\end{figure}

The final step is to gather texels from the lanes in the wave and filter them.
\chreplaced{For each two-dimensional texel coordinate needed in a lane, we compute the
corresponding integer coordinate, $t$, and then retrieve the desired texel from lane}%
{Each lane knows which texels in the mask are required for its filter.
Let the upper left coordinate of the lane's $2\times 2$ region
be $t\in[0,255]$. All the
$2\times 2$ texel coordinates (1D) are then $\{t, t+1, t+16, t+16+1\}$, where $16$
is the $16\times 16$ mask width. The goal is now
to go from texture coordinate, e.g., from $t$ to lane index, $i$,
and retrieve the texel that lane $i$ has \chreplaced{produced}{looked up} there.
We do this by} $i = h^{-1}(t, B)$, where the inverse of
finding the $i$th set bit is counting the set bits below bit $t$. 
This is illustrated in Figure~\ref{fig_bijective_function_h}.
\chreplaced{All the}{The} \chdeleted{$2\times 2$} texels needed for the current lane are guaranteed to have been produced
by \chdeleted{four of the lanes in }the wave, and these are weighted and
the perfectly-filtered value is returned.

A $16\times 16$ bitmask is sufficient for both bilinear and bicubic filters,
though a smaller bitmask may be used to reduce register pressure and computation,
potentially at a cost in effectiveness.
We have found empirically that an $11\times 11$ bitmask stored in two \texttt{uint64\_t} variables
is sufficient for the bilinear filter---it gives identical results to the $16\times 16$
bitmask with 25--33\% lower runtime \chadded{in our experiments}. %
We use this variant for measurements in Section~\ref{sec_results}.

\subsection{New Fallback Methods}	%
\label{sec_fallback_method}
\chreplaced{This section presents novel fallback methods that may be used to handle the case when
perfect filtering cannot be achieved (discussed in Section~\ref{sec_results}).
In these methods, we attempt to make each lane in the wave do useful work toward increasing the filtering quality.}
{While our algorithms are often able to produce perfect filtering,
the degree of magnification is sometimes too low to be able to produce all the required texels with one texture evaluation per lane.
Such cases are illustrated by the colored regions in Figure~\ref{fig_lookups} for the bilinear filter.
(See Figure~\ref{fig:bicubic_lookups} in our supplemental material for the bicubic counterpart.)
Those occur only for low magnification factors.
Both \WL and \WM produce flawless results when the magnification factor is $>1.59$, for example.
\WB is more conservative and requires higher magnification factors ($>2.35$)
to achieve perfect results.
We present novel fallback methods that may be used to handle
the case when perfect bilinear filtering cannot
be achieved. They attempt to make each lane in the wave
do useful work toward increasing the filtering quality.}

In our fallback methods,
each lane starts by computing the coordinates of a texel using STF,
i.e., randomly selecting \chreplaced{a}{the} texel with probability based on its corresponding filter weight.
In our simplest fallback variant, all lanes \chreplaced{produce}{perform} their
corresponding STF texel \chdeleted{lookup}, gather unique texels within the wave belonging to their filter footprints and then weight those texels\chdeleted{,
as described by Equation~\ref{eq_fallback},} to form the final filtered result.
Assume that a texel value, $\mathbf{p}_k$, is produced for lane $k$.
Each lane $i$ identifies $N$ unique texels $\mathbf{p}_i$,
with nonzero filter weights $w_i$,
that contribute to its pixel's filtered value.
\chdeleted{We note that this number is guaranteed to be at least one, and at most equal to the size of the texture filter footprint, e.g., four for the bilinear filter.}
Then, we estimate each lane's filtered value as  %
\begin{equation}
	\mathbf{c}_k =
	\sum_{i=0}^{N-1} w_i \mathbf{p}_i + \left(1 - \sum_{i=0}^{N-1} w_i\right) \frac{\sum_{i=0}^{N-1} \mathbf{p}_i}{N},
	\label{eq_fallback}
\end{equation}
where the first term represents the known filtered texel values weighted by their corresponding filter weights.
The second term is an estimate of the missing texel values as an unweighted average of the known ones.
This estimate would be unbiased if it were not for the lane PDF mismatch~\cite{Wronski2025}.
For the case when we only have one unique texel for lane $i$,
we return a single value $\mathbf{p}_i$, the same as
the classic \STF \chreplaced{evaluation}{lookup}~\cite{Pharr2024}.
When all texel values necessary for filtering a single lane are known and present in the wave,
the second term of the equation is zero and we get perfect \chdeleted{and deterministic} filtering with zero error.

An extension to the method above further improves quality.
We note that just like in the method described by Wronski et al.~\cite{Wronski2025},
multiple lanes may plan to produce the same texel, resulting in redundant work.
To remove some of them, we \chreplaced{represent}{identify} the \chdeleted{shared} set of texels \chadded{to be produced} using a
bitmask,
similar to \WM.
Before the actual STF texel \chreplaced{evaluation}{lookup},
each lane sets the bit
corresponding to the coordinates of its STF texel. We then use
\texttt{WaveActiveBitOr()} to find a bitmask containing all
texels that the wave is set to produce. Counting the number $n$ of set bits in that mask
gives us the number of unique texels that need to be produced for
this initial texel \chreplaced{evaluation}{lookup}. The difference here compared to \WM
is that we only set one bit per lane, rather than $N\times N$.
We can then
leverage the remaining $32-n$ lanes to improve quality.
The first $n$ lanes \chadded{each} \chreplaced{produce}{perform} their
corresponding STF texel\chdeleted{ lookup}. 
The remaining $32-n$ lanes also produce texels, but we 
let each of these stochastically select a texel from its filter footprint
that has not yet been \chreplaced{produced}{looked up} by any of the first $n$ lanes.
These \chreplaced{evaluations}{lookups} are spread out evenly over the wave's 32 lanes. %
When $n<32$, this is done with a simple mapping, i.e., 
\begin{equation}
	l = \left\lfloor \frac{31(c-n)}{31-n}\right\rceil,
\end{equation}
where $l$ is the resulting lane number, $c$ is the current lane number, %
which in the case of the $32-n$ remaining lanes must be in $[n, 31]$,
and $\lfloor\cdot\rceil$ rounds to the nearest integer.
Our mapping ensures that lanes 0 and 31 always are included when
$32-n \geq 2$. When only one lane is left, the equation above maps it to $l=0$.
Alternative mappings are also possible.
At this point all lanes in the wave have produced a texel,
and then weighting is again done as described by Equation~\ref{eq_fallback}.

\chreplaced{These}{Our suggested} fallback methods are biased\chreplaced{, yet}{. Yet,} as shown
in Section~\ref{sec_results}, \chreplaced{they}{these new methods} provide
superior quality compared to previous, unbiased alternatives.
In the results section, we mark the use of the simpler fallback method described above
with \textbf{C}, and the use of the extended version with \textbf{C+}.

\section{Results}
\label{sec_results}

We start by focusing on bilinear filtering for magnification,
with bicubic filtering evaluated in Section~\ref{sec_results_bicubic}.
Additional performance considerations are discussed in Section~\ref{sec_perf_expensive_decompressors}.
For visual results, we refer to Figure~\ref{fig_teaser}, Figure~\ref{fig:bicubic},
and our supplemental video.

\chadded{All methods are implemented in the Falcor rendering
framework~\cite{Kallweit2022}.} %
\chadded{
To measure quality \textit{only} on waves with magnification, we use
trilinear texture filtering for all waves that have at least one pixel with minification.}
\chreplaced{Performance is}{The performance results are} measured on an NVIDIA RTX 5090 GPU.
Quality results are aggregated over five different textures; see Section~\ref{sec:scenes}
for images and discussion of how image metrics were aggregated.
Furthermore, as denoising can
be expected to be applied as a post-process to most rendering algorithms,
image sequences are denoised using the DLSS~\cite{NVIDIA2025:DLSS} spatiotemporal denoiser unless stated otherwise.
We use its \texttt{DLAA} mode,
which does spatiotemporal denoising and antialiasing, but no superresolution.
\chreplaced{Diagrams generated}{Figures} with non-denoised images corresponding to ones presented in this section
can be found in our supplemental material.
Our runtime performance measurements focus on the incremental costs of our algorithms and do not include DLSS,
which we assume is already in use.

We primarily use the ColorVideoVDP video quality metric~\cite{Mantiuk2024} (CVVDP for
short) to evaluate error; it attempts to model spatiotemporal aspects of human vision to
better correspond to human judgment than simpler metrics such as PSNR.
ColorVideoVDP's output is in just-objectionable differences (JODs). The maximum
JOD value is 10, which corresponds to the image or video being
perceptually indistinguishable from the ground truth.
Table~\ref{tab:results_table_combined}
also includes results with other image metrics as well as error measurements of non-denoised sequences.

\subsection{Bilinear Filtering}
\label{sec_results_bilinear}
\begin{figure}[t]
	\centering
	\includegraphics[width=\columnwidth]{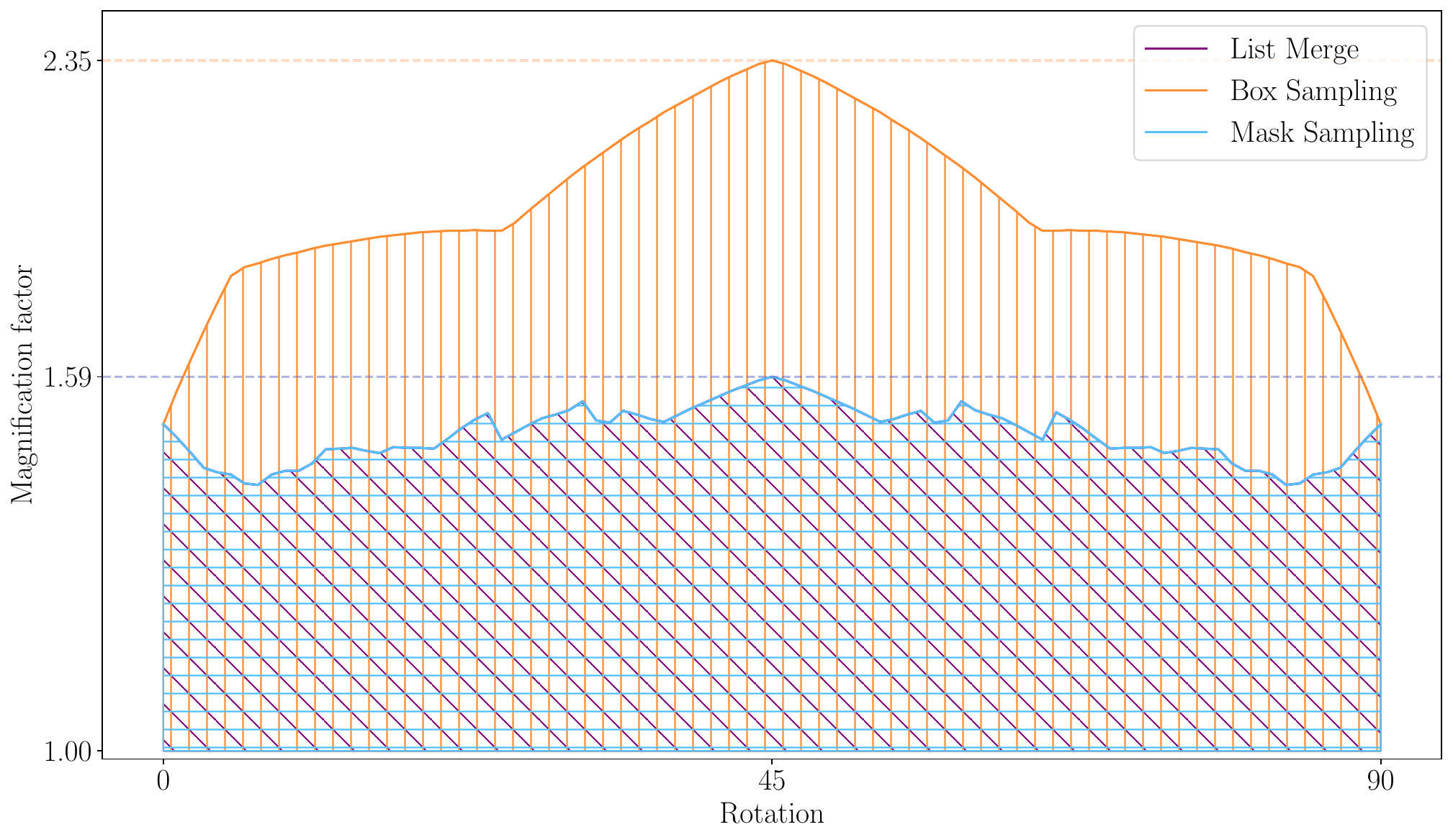}
	\caption{Plot of magnification factors where our algorithms
		are unable to produce all texels necessary for perfect bilinear filtering
		and need to rely on fallback methods (Section~\ref{sec_fallback_method}).
		This magnification factor is a function of
		the rotation of the texel grid compared to the screen pixel grid.
		The scene used was a rotated, textured quad, with the camera
		set to view the quad's center head on.
		Cases where a fallback was necessary are indicated by colored areas.
		\chdeleted{We can see that }\WL and \WM are the most effective, achieving perfect filtering
		around a magnification factor of 1.59 \chadded{and above} for all quad orientations.
		(Note that both cover identical areas.)
		The success rate of \WB varies more with orientation, with a magnification factor
		of \chadded{at least} 2.35 needed for perfect filtering at the challenging case of a 45 degree rotation.
	}
	\label{fig_lookups}
\end{figure}

We start with results for bilinear texture filtering during magnification,
which are used to support the recommendations we make in Section~\ref{sec_discussion}.
\chadded{
While our algorithms are often able to produce perfect filtering,
the degree of magnification is sometimes too low to be able to produce all the required texels with one texture evaluation per lane.
In such cases, our fallback methods (Section~\ref{sec_fallback_method}) are needed.
For the bilinear filter, those cases
are illustrated by the colored regions in Figure~\ref{fig_lookups}.
See Figure~\ref{fig:bicubic_lookups} in our supplemental material for the bicubic counterpart.
}

\begin{figure}[h!]
	\setlength{\belowcaptionskip}{-35pt} %
	\begin{minipage}[t]{\columnwidth}
		\centering
		\includegraphics[width=\linewidth]{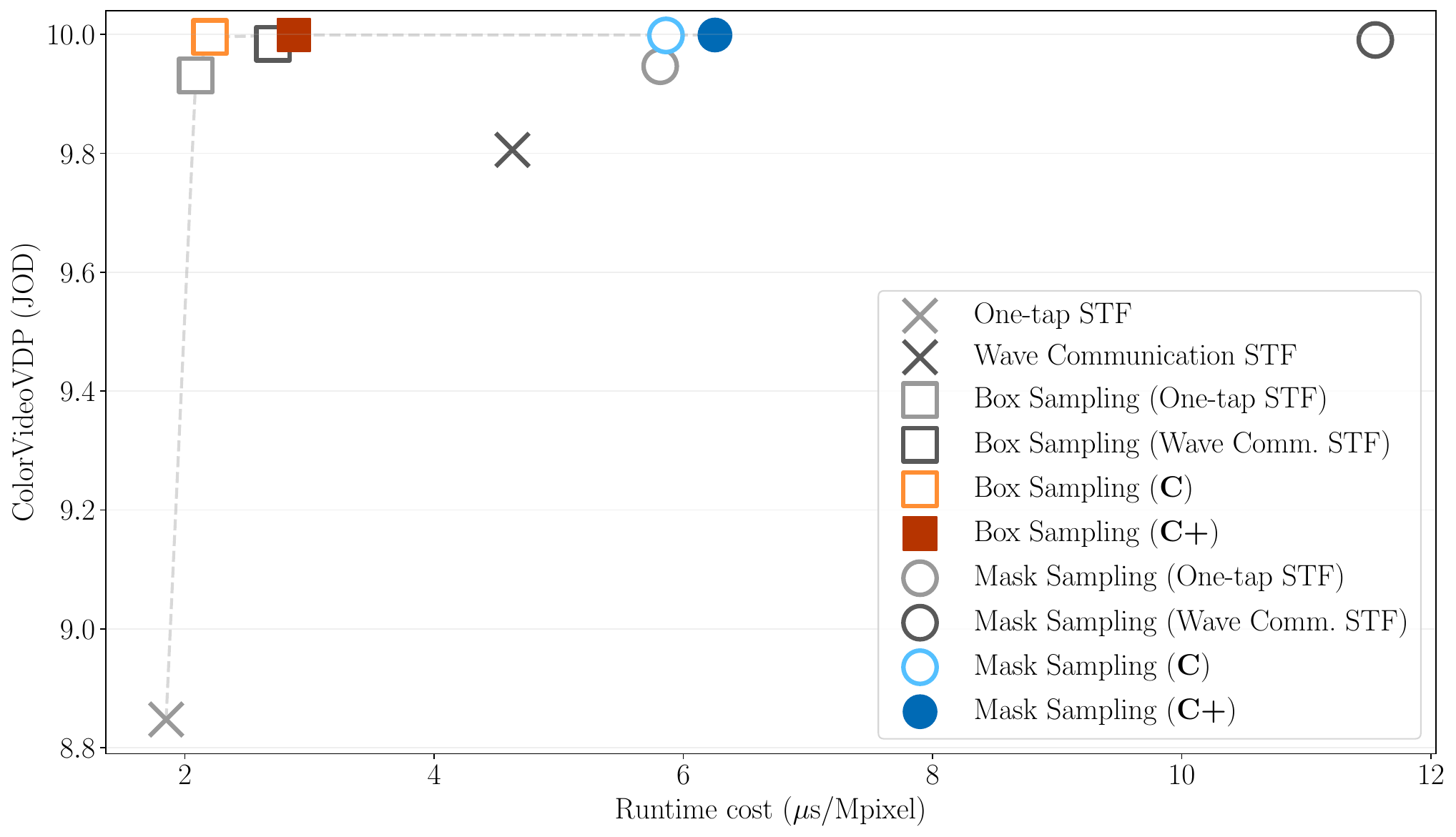}
		\caption{Pareto frontier (dashed line) indicating the most quality/performance-efficient
			algorithm alternatives.
			The corresponding PSNR range for this plot was approximately \chhighlight{35--67}\chdeleted{33--58}~dB.}
		\label{fig:paretoDLSS}
	\end{minipage}
	\\

	\vspace{5.45em}

	\begin{minipage}[t]{\columnwidth}
		\centering
		\includegraphics[width=\linewidth]{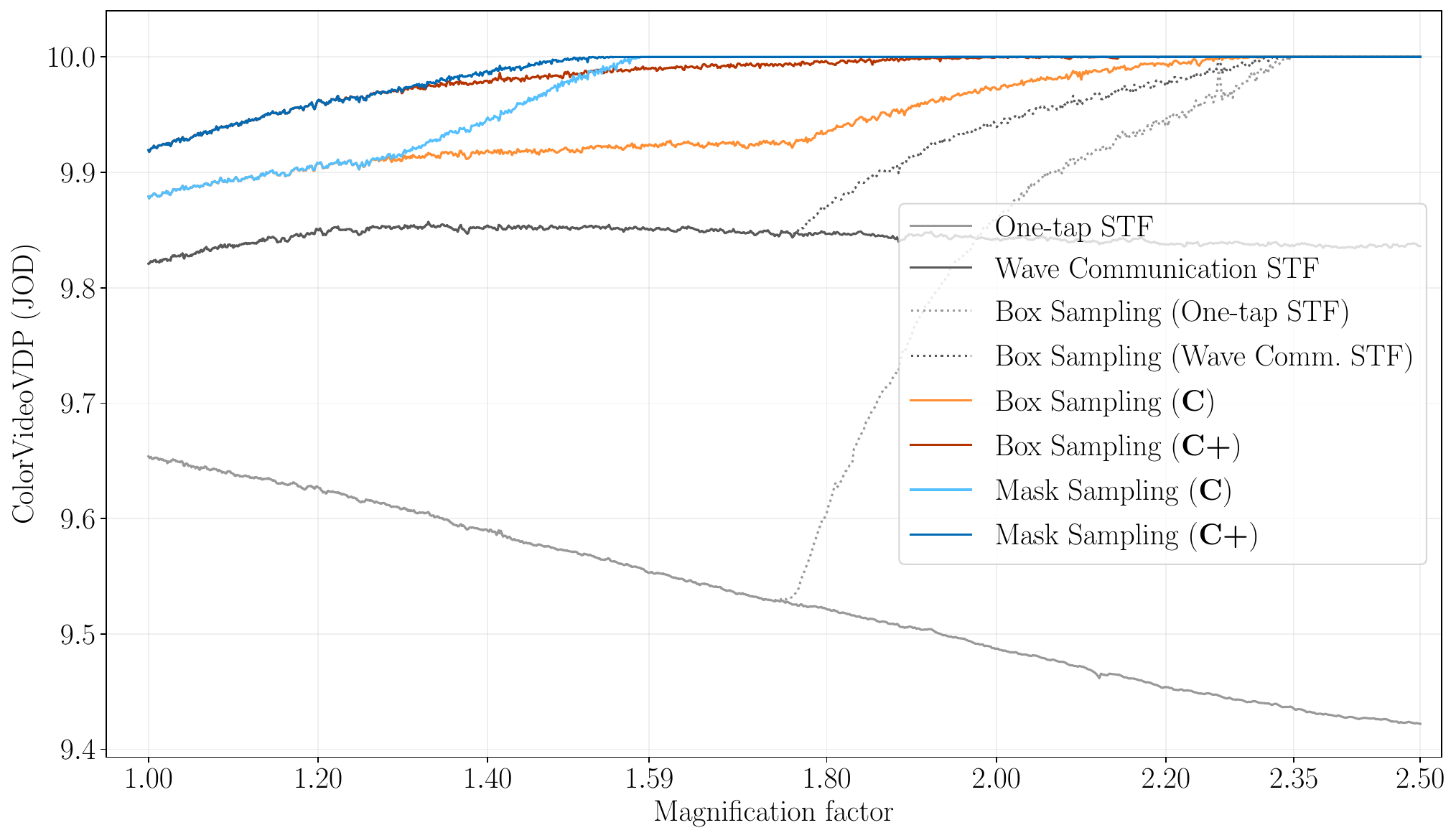}
		\caption{Quality at different magnification factors. As indicated by Figure~\ref{fig_lookups},
			perfect bilinear filtering is achieved for magnification factors above $2.35$
			for all the methods we propose. \WM achieves perfect filtering at lower magnification factors
			than \WB. STF quality declines as magnification increases due to increased aliasing and error
			from filtering after shading~\cite{Pharr2024}.}
		\label{fig:quality_zoom}
	\end{minipage}
	\\

	\vspace{5.45em}

	\begin{minipage}[t]{\columnwidth}
		\centering
		\includegraphics[width=\linewidth]{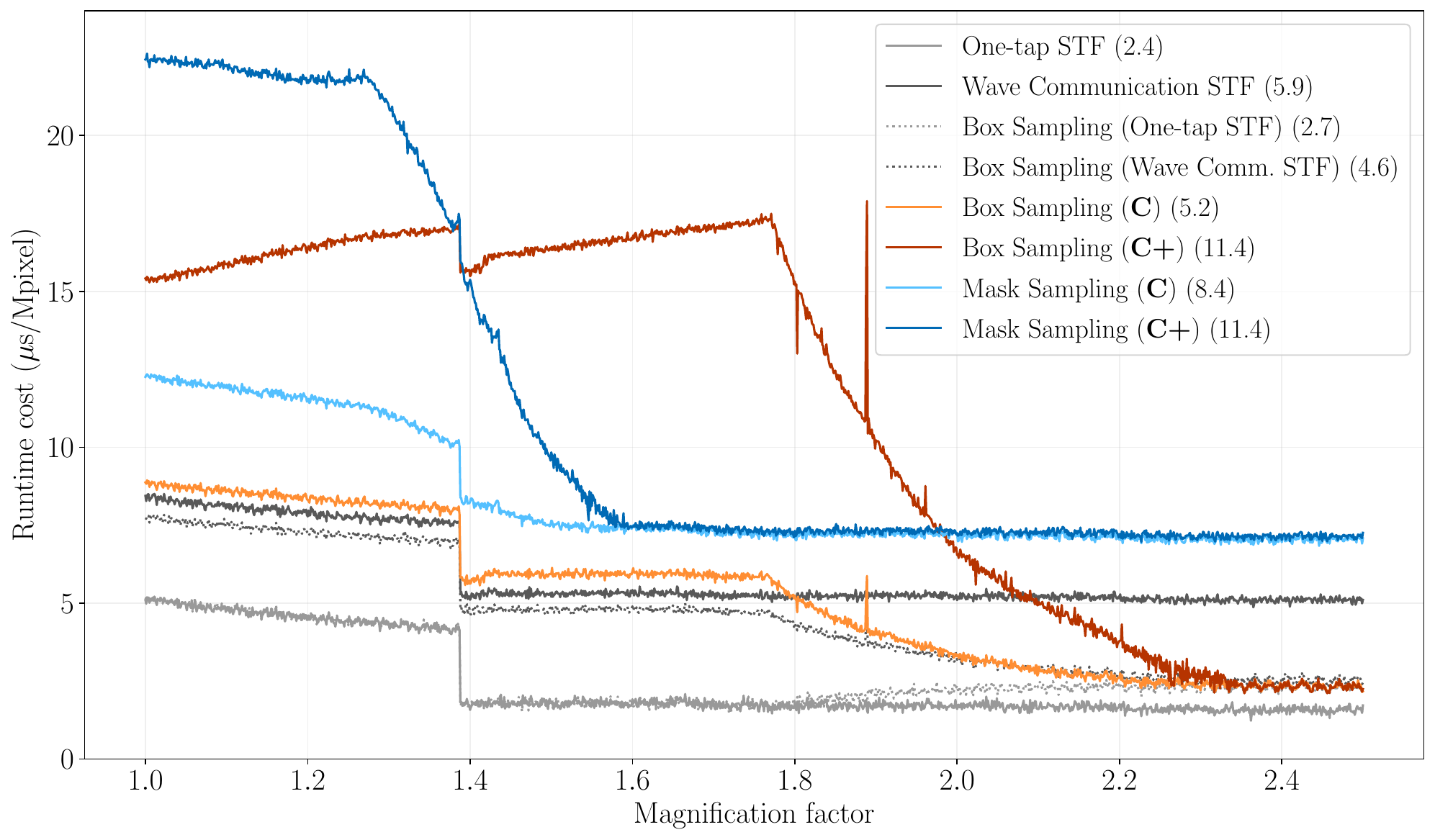}
		\caption{Performance for an NVIDIA RTX 5090
			under different levels of magnification.
			The numbers in parentheses show the algorithms' average runtime costs.
			For reference, running at 60~FPS at $2560\times 1440$ corresponds to
			4741 $\mu$s/Mpixel, indicating that all methods consume a tiny fraction of total frame time.}
		\label{fig:performance_5090}
	\end{minipage}
\end{figure}

Figure~\ref{fig:paretoDLSS} plots the quality and performance of our new methods
and the state-of-the-art alternatives---\STF~\cite{Pharr2024} and \WC~\cite{Wronski2025}
both as standalone methods but also as fallback options for Box and \WM.
Figures~\ref{fig:quality_zoom} and \ref{fig:performance_5090},
discussed later in the section, give
more detailed presentations of the algorithms' quality and performance. %
\WL was not considered because \WM reaches the same success rate
of achieving perfect filtering (Figure~\ref{fig_lookups}) while
incurring a smaller runtime cost. 
The diagrams also contain combinations of our proposed fallback methods
with Box and \WM as main methods. 
In all of our results, the
fallback method used is indicated in parenthesis, e.g., ``\WM (\textbf{C+})''
is \WM with our \textbf{C+} technique as fallback. This combination
is what we call ``Ours'' in Figure~\ref{fig_teaser}.
The scenes used for our evaluation are described in our supplemental material.
The one used for Figure~\ref{fig:paretoDLSS} and Table~\ref{tab:results_table_combined}
contains a range of magnification factors to make it similar to a realistic use case.

Considering Figure~\ref{fig:paretoDLSS} and Table~\ref{tab:results_table_combined}, %
we see that \chreplaced{our}{the combinations of our main and fallback} methods provide higher-quality results
than \STF and \WC, though \STF
is the fastest method.
Following the Pareto frontier in Figure~\ref{fig:paretoDLSS}, we see that combining \WB
with \STF \chadded{as a fallback} gives a large increase in quality compared to
\STF by itself without significantly increasing runtime cost.
Our fallback methods may then increase quality further,
though with a small performance cost.
\WM paired with \STF and \WC \chadded{fallbacks}
are not quality/performance-efficient options.
Those combinations are therefore excluded in the remainder of the paper.
\chdeleted{We note that the same methods are Pareto-optimal and the Pareto frontier
has the same shape when PSNR is used as the image quality metric.}

In Table~\ref{tab:results_table_combined},
we complement the quality results presented in Figure~\ref{fig:paretoDLSS}
with the PSNR and \FLIP~\cite{Andersson2020}
image quality metrics, as well as measurements for non-denoised sequences.
Our methods give significantly higher quality than
the state-of-the-art algorithms.
Denoising reduces the quality gap between our and prior
techniques, which are designed to be used with denoising.

In order to analyze quality and performance as magnification varies,
we next consider a scene where the camera is
set up to view a rotated quad's center head on. \chreplaced{Given the results in}{Based on} Figure~\ref{fig_lookups},
we rotated the quad 45 degrees, as that is the most challenging
case. Figure~\ref{fig:quality_zoom}
shows how quality varies as magnification increases.
\WM achieves perfect \chdeleted{bilinear} filtering at lower
levels of magnification than \WB while previous algorithms
do not achieve this for any level of magnification.
For additional quality results, including convergence rates and
maximum errors, see Section~\ref{supp_results} in our supplemental material.

Figure~\ref{fig:performance_5090} shows how the runtime cost of the methods
changes during the same animation using a single lookup with a standard RGBA texture. %
We can see that
our extended fallback method (\textbf{C+}) contributes a large portion of the total runtime cost
for the variants that use it.
This is likely due to its implementation requiring more registers than
\chdeleted{the} simpler alternatives.
As magnification increases, the less our algorithms need to resort to fallback methods,
\chreplaced{which reduces}{eventually reducing} their runtime cost. %

\subsection{Bicubic Filtering}\label{sec_results_bicubic}
Our comparisons so far have focused on the bilinear filter due to its ubiquity in real-time rendering.
However, our methods also work with arbitrary discrete filters.
Figure~\ref{fig:bicubic} shows visual results and error measurements using a bicubic B-spline filter
as well as the Catmull--Rom filter.
The results parallel those in Figure~\ref{fig_teaser}, with our algorithm giving
significantly higher quality than previous approaches.

The Catmull--Rom filter includes negative weights. To sample such filters,
\STF uses a Monte Carlo technique called positivization~\cite{Pharr2024}.
Positivization requires sampling the positive \chreplaced{\textit{and}}{and} negative lobes of the filter separately,
thus requiring \chreplaced{that}{producing} two texels per lane \chadded{are produced}.
Our method not only provides much higher quality (perfect filtering with zero error for sufficient\chadded{ly large} magnification \chadded{factors}), but also
reduces the number of texels produced to $\leq 1$ per lane.
\chreplaced{We}{Furthermore, we} make no distinction between positive and negative weights and the types of filters\chreplaced{ other than}{.
No code modifications are needed, except} computing absolute values of filter weights used for sampling probabilities in our fallback method. %

\subsection{Performance with Expensive Texture Decompression}
\label{sec_perf_expensive_decompressors}

\chadded{
In this section, we first present performance results
using an existing neural texture decompressor and
then consider DCT decompression, where our techniques can be
applied to further reduce the cost of texel evaluation by
distributing work for a single texel across multiple lanes.
}
\subsubsection{\chadded{Neural Texture Decompression}}
\chadded{
To evaluate how our algorithms perform with
neural texture compression~\cite{Vaidyanathan2023} (NTC),
we 
added \WB with \STF as fallback into the \texttt{ntc-renderer}
in the NTC SDK.}\footnote{\chadded{\url{https://github.com/NVIDIA-RTX/RTXNTC}}}
\chadded{We rendered the scene \textit{FlightHelmet} and zoomed in on a part of the helmet
such that the magnification factor was $>2.35$ for every pixel.
All NTC evaluations
used \textit{inference on sample}, i.e., texels were decompressed when needed
by the renderer.
The renderer ran at 7.7~$\mu$s/Mpixel without any texel evaluation,
and with one NTC evaluation per pixel using \STF, it used 89~$\mu$s/Mpixel.
Our \WB with \STF as fallback ran at 93~$\mu$s/Mpixel.
(Recall also that rendering at, for example, $2560\times 1440$ pixels at 60~FPS
is equivalent to $4741~\mu$s/Mpixel.)
Due to the level of magnification, our method produced an image with zero error,
while \STF generated noisy images.
We also implemented full bilinear filtering where $2\times 2$ NTC texels
were decompressed for each pixel. Performance was $516$~$\mu$s/Mpixel
with the same image quality as our method but more than 
$4\times$ higher runtime.
}

\begin{table*}[t]
\centering
\footnotesize
\definecolor{lightgray}{rgb}{0.9,0.9,0.9}
\captionof{table}{Error metrics averaged across sequences. $\downarrow$ indicates lower is better, $\uparrow$ indicates higher is better. For each entry, the first number (black) shows results without denoising while the second number (non-black) shows results with denoising.}
\label{tab:results_table_combined}
\resizebox{\textwidth}{!}{
\begin{tabular}{l @{\hspace{0.9em}}c @{\hspace{0.9em}}c @{\hspace{0.9em}}c @{\hspace{0.9em}}c @{\hspace{0.9em}}c @{\hspace{0.9em}}c @{\hspace{0.9em}}c @{\hspace{0.9em}}c}
\toprule
& \makecell{One-tap STF\\\cite{Pharr2024}}& \makecell{Wave Comm. STF\\\cite{Wronski2025}}& \makecell{Box Sampling\\(One-tap STF)}& \makecell{Box Sampling\\(Wave Comm.)}& \makecell{Box Sampling\\(\textbf{C})}& \makecell{Mask Sampling\\(\textbf{C})}& \makecell{Box Sampling\\(\textbf{C+})}& \makecell{Mask Sampling\\(\textbf{C+})} \\
\midrule
PSNR ($\uparrow$) & 28.57 \textcolor{lightgray}{|} \textcolor{rgb,1:red,0.6;green,0.6;blue,0.6}{34.94} & 34.93 \textcolor{lightgray}{|} \textcolor{rgb,1:red,0.35;green,0.35;blue,0.35}{42.60} & 42.92 \textcolor{lightgray}{|} \textcolor{rgb,1:red,0.6;green,0.6;blue,0.6}{52.15} & 48.04 \textcolor{lightgray}{|} \textcolor{rgb,1:red,0.35;green,0.35;blue,0.35}{56.63} & 51.99 \textcolor{lightgray}{|} \textcolor{rgb,1:red,1.0;green,0.5583632666810756;blue,0.19818011428336807}{60.06} & 58.02 \textcolor{lightgray}{|} \textcolor{rgb,1:red,0.3277250527283637;green,0.7533866722110684;blue,1.0}{65.50} & 58.40 \textcolor{lightgray}{|} \textcolor{rgb,1:red,0.7127015300411498;green,0.204225299979703;blue,0.0}{65.56} & \textbf{59.48} \textcolor{lightgray}{|} \textcolor{rgb,1:red,0.0;green,0.41394354779309744;blue,0.7090815962463488}{\textbf{66.58}} \\
\arrayrulecolor{lightgray}\midrule\arrayrulecolor{black}
ColorVideoVDP ($\uparrow$) & 8.792 \textcolor{lightgray}{|} \textcolor{rgb,1:red,0.6;green,0.6;blue,0.6}{8.848} & 9.618 \textcolor{lightgray}{|} \textcolor{rgb,1:red,0.35;green,0.35;blue,0.35}{9.806} & 9.887 \textcolor{lightgray}{|} \textcolor{rgb,1:red,0.6;green,0.6;blue,0.6}{9.931} & 9.941 \textcolor{lightgray}{|} \textcolor{rgb,1:red,0.35;green,0.35;blue,0.35}{9.985} & 9.973 \textcolor{lightgray}{|} \textcolor{rgb,1:red,1.0;green,0.5583632666810756;blue,0.19818011428336807}{9.996} & 9.987 \textcolor{lightgray}{|} \textcolor{rgb,1:red,0.3277250527283637;green,0.7533866722110684;blue,1.0}{9.998} & 9.989 \textcolor{lightgray}{|} \textcolor{rgb,1:red,0.7127015300411498;green,0.204225299979703;blue,0.0}{\textbf{9.999}} & \textbf{9.990} \textcolor{lightgray}{|} \textcolor{rgb,1:red,0.0;green,0.41394354779309744;blue,0.7090815962463488}{\textbf{9.999}} \\
\arrayrulecolor{lightgray}\midrule\arrayrulecolor{black}
\FLIP ($\downarrow$) & 0.0501 \textcolor{lightgray}{|} \textcolor{rgb,1:red,0.6;green,0.6;blue,0.6}{0.0328} & 0.0375 \textcolor{lightgray}{|} \textcolor{rgb,1:red,0.35;green,0.35;blue,0.35}{0.0224} & 0.0034 \textcolor{lightgray}{|} \textcolor{rgb,1:red,0.6;green,0.6;blue,0.6}{0.0040} & 0.0028 \textcolor{lightgray}{|} \textcolor{rgb,1:red,0.35;green,0.35;blue,0.35}{0.0036} & 0.0018 \textcolor{lightgray}{|} \textcolor{rgb,1:red,1.0;green,0.5583632666810756;blue,0.19818011428336807}{0.0027} & 0.0003 \textcolor{lightgray}{|} \textcolor{rgb,1:red,0.3277250527283637;green,0.7533866722110684;blue,1.0}{0.0016} & 0.0006 \textcolor{lightgray}{|} \textcolor{rgb,1:red,0.7127015300411498;green,0.204225299979703;blue,0.0}{0.0018} & \textbf{0.0002} \textcolor{lightgray}{|} \textcolor{rgb,1:red,0.0;green,0.41394354779309744;blue,0.7090815962463488}{\textbf{0.0015}} \\
\bottomrule
\end{tabular}
}
\end{table*}

\begin{figure*}[t!]
	\setlength{\tabcolsep}{0.75pt}
	\renewcommand{\arraystretch}{0.5}	
	\newcommand{\mywidth}{11.15mm}
	\newcommand{\outerLineWidth}{0.9pt}
	\newcommand{\innerLineWidth}{0.4pt}
	\newcommand{\grayLevel}{75}
	\begin{tabular}{ccc!{\color{gray!\grayLevel}\vrule width \outerLineWidth}cc!{\color{gray!\grayLevel}\vrule width \outerLineWidth}cc!{\color{gray!\grayLevel}\vrule width \outerLineWidth}cc!{\color{gray!\grayLevel}\vrule width \outerLineWidth}!{\color{gray!\grayLevel}\vrule width \outerLineWidth}cc!{\color{gray!\grayLevel}\vrule width \outerLineWidth}cc!{\color{gray!\grayLevel}\vrule width \outerLineWidth}cc}
		&		
		\multicolumn{8}{c!{\color{gray!\grayLevel}\vrule width \outerLineWidth}!{\color{gray!\grayLevel}\vrule width \outerLineWidth}}{\textbf{Bicubic B-spline}} &
		\multicolumn{6}{c}{\textbf{Bicubic Catmull--Rom}} 
		\\
		&
		\multicolumn{2}{c!{\color{gray!\grayLevel}\vrule width \outerLineWidth}}{Ground truth} &
		\multicolumn{2}{c!{\color{gray!\grayLevel}\vrule width \outerLineWidth}}{\STF} &
		\multicolumn{2}{c!{\color{gray!\grayLevel}\vrule width \outerLineWidth}}{Wave Comm.\ STF} &
		\multicolumn{2}{c!{\color{gray!\grayLevel}\vrule width \outerLineWidth}!{\color{gray!\grayLevel}\vrule width \outerLineWidth}}{Mask Sampl.\ (\textbf{C})} &
		\multicolumn{2}{c!{\color{gray!\grayLevel}\vrule width \outerLineWidth}}{Ground truth} &
		\multicolumn{2}{c!{\color{gray!\grayLevel}\vrule width \outerLineWidth}}{\STF} &
		\multicolumn{2}{c}{Mask Sampl.\ (\textbf{C})}
		\\
		& 		
		\multicolumn{2}{c!{\color{gray!\grayLevel}\vrule width \outerLineWidth}}{{\scriptsize PSNR / \FLIP~/ CVVDP}} &
		\multicolumn{2}{c!{\color{gray!\grayLevel}\vrule width \outerLineWidth}}{{\footnotesize 26.7 / 0.057 / 8.83}} &
		\multicolumn{2}{c!{\color{gray!\grayLevel}\vrule width \outerLineWidth}}{{\footnotesize 34.5 / 0.043 / 9.24}} &
		\multicolumn{2}{c!{\color{gray!\grayLevel}\vrule width \outerLineWidth}!{\color{gray!\grayLevel}\vrule width \outerLineWidth}}{\textbf{{\footnotesize 48.0 / 0.0071 / 9.87}}} &
		\multicolumn{2}{c!{\color{gray!\grayLevel}\vrule width \outerLineWidth}}{{\scriptsize PSNR / \FLIP~/ CVVDP}} &
		\multicolumn{2}{c!{\color{gray!\grayLevel}\vrule width \outerLineWidth}}{{\footnotesize 27.3 / 0.060 / 8.81}} &
		\multicolumn{2}{c}{\textbf{{\footnotesize 46.9 / 0.0064 / 9.88}}}
		\\
		\rotatebox{90}{{\scriptsize not denoised}} &
		\includegraphics[width=\mywidth,cfbox=red 1pt 0pt]{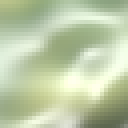} & 
		\includegraphics[width=\mywidth,cfbox=yellow 1pt 0pt]{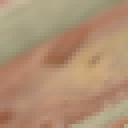} &
		\includegraphics[width=\mywidth]{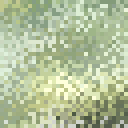} & 
		\includegraphics[width=\mywidth]{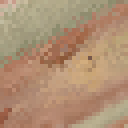} &
		\includegraphics[width=\mywidth]{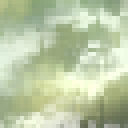} & 
		\includegraphics[width=\mywidth]{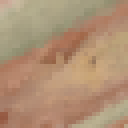} &
		\includegraphics[width=\mywidth]{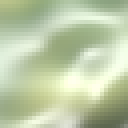} & 
		\includegraphics[width=\mywidth]{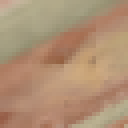} &
		\includegraphics[width=\mywidth,cfbox=red 1pt 0pt]{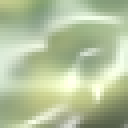} &
		\includegraphics[width=\mywidth,cfbox=yellow 1pt 0pt]{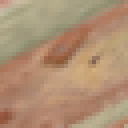} &
		\includegraphics[width=\mywidth]{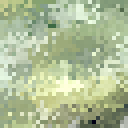} &		
		\includegraphics[width=\mywidth]{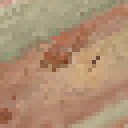} &		
		\includegraphics[width=\mywidth]{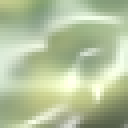} &	
		\includegraphics[width=\mywidth]{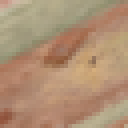} 		
		\\
		\rotatebox{90}{{\scriptsize denoised}} &
		\includegraphics[width=\mywidth,cfbox=red 1pt 0pt]{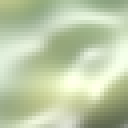} & 
		\includegraphics[width=\mywidth,cfbox=yellow 1pt 0pt]{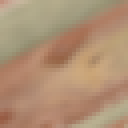} &
		\includegraphics[width=\mywidth]{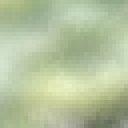} & 
		\includegraphics[width=\mywidth]{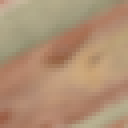} &
		\includegraphics[width=\mywidth]{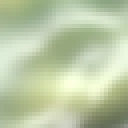} & 
		\includegraphics[width=\mywidth]{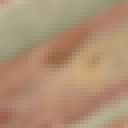} &
		\includegraphics[width=\mywidth]{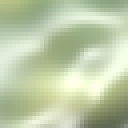} & 
		\includegraphics[width=\mywidth]{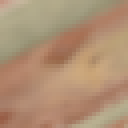} &
		\includegraphics[width=\mywidth,cfbox=red 1pt 0pt]{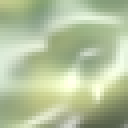} &
		\includegraphics[width=\mywidth,cfbox=yellow 1pt 0pt]{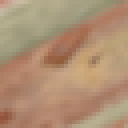} &
		\includegraphics[width=\mywidth]{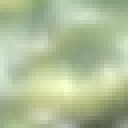} &		
		\includegraphics[width=\mywidth]{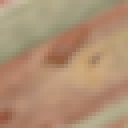} &		
		\includegraphics[width=\mywidth]{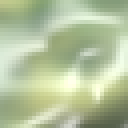} &	
		\includegraphics[width=\mywidth]{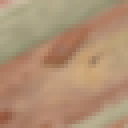} \\		
		&
		\multicolumn{2}{c!{\color{gray!\grayLevel}\vrule width \outerLineWidth}}{{\scriptsize PSNR / \FLIP~/ CVVDP}} &
		\multicolumn{2}{c!{\color{gray!\grayLevel}\vrule width \outerLineWidth}}{{\footnotesize 35.5 / 0.044 / 9.06}} &
		\multicolumn{2}{c!{\color{gray!\grayLevel}\vrule width \outerLineWidth}}{{\footnotesize 44.5 / 0.026 / 9.80}} &
		\multicolumn{2}{c!{\color{gray!\grayLevel}\vrule width \outerLineWidth}!{\color{gray!\grayLevel}\vrule width \outerLineWidth}}{\textbf{{\footnotesize 55.0 / 0.0062 / 9.96}}} &
		\multicolumn{2}{c!{\color{gray!\grayLevel}\vrule width \outerLineWidth}}{{\scriptsize PSNR / \FLIP~/ CVVDP}} &
		\multicolumn{2}{c!{\color{gray!\grayLevel}\vrule width \outerLineWidth}}{{\footnotesize 36.1 / 0.044 / 9.23}} &
		\multicolumn{2}{c}{\textbf{{\footnotesize 53.3 / 0.0061 / 9.95}}}
	\end{tabular}
	\caption{Visual comparison and error measurments with the bicubic B-spline and Catmull--Rom filters.
		All three error metrics indicate that our method achieves very high quality, both with and without denoising.
		For Catmull--Rom filters with negative weights, our method requires \chhighlight{$0.88$}\chdeleted{$0.74$} \chreplaced{texel evaluations}{lookups} per pixel,
		while \STF requires two \chreplaced{texel evaluations}{texture lookups} to sample the positive and negative lobes of the filter separately.
		\WC does not support filters with negative weights. %
	}
	\label{fig:bicubic}
\end{figure*}
\subsubsection{\chadded{DCT Decompression}}
\label{sec_dct_decompression}
When there are more active lanes in a wave than needed texels,
there is an opportunity to distribute the work required for producing a texel value across multiple lanes
rather than having a single lane produce each texel value;
this can give \chreplaced{improved}{better} performance\chdeleted{ with compressed textures that are expensive to decompress}.
As a proof-of-concept, we implemented a simple block-based discrete cosine transform (DCT) compressed texture representation
where decompress\chreplaced{ing a block}{ion} requires $8 \times 8$ matrix multiplications.

For a baseline, we decompress each texel in a single lane, first decoding the red channel, then green and blue.
We then added an optimized path for \chadded{when} ten or fewer texels \chadded{are} needed in a 32-lane wave---in that case, lanes decompress just
a single color channel and color channels are then assembled into full \chadded{RGB} texel values via wave intrinsics.
With a test scene with a plane viewed at an angle where this optimization was applicable at roughly half of the pixels, we saw a $1.64\times$ speedup.
When all pixels had sufficient\chadded{ly large} magnification \chadded{factors}, we measured a $2.5\times$ speedup.
Similar performance improvements may be available for neural compression
techniques~\cite{Vaidyanathan2023,Fujieda2024,Kim2024neuralvdb,Dupuy2025},
where neural network evaluation could be distributed across multiple lanes when idle ones are available.
However, a full performance evaluation is out of the scope of this
paper and left for future work.

\chdeleted{
To put the performance numbers presented in Figure~\ref{fig:performance_5090}
into perspective, we compare them to the cost of decompressing neural textures.
On an RTX 5090, NTC decompression when textures are loaded into memory (\emph{inference on load}) can achieve
14~Gpixels/s}\chdeleted{, or $71~\mu$s/Mpixel.
We would expect greater overhead for \emph{inference on sample} where decompression is
performed in shaders during rendering as would be done using our methods.
As shown in Figure~\ref{fig:performance_5090},
our methods use from $5.2~\mu$s/Mpixel (\WB (\textbf{C}))
up to $11.4~\mu$s/Mpixel (\WM (\textbf{C+})) and so use much less time than NTC decompression of a single texel.
(Recall also that rendering at, for example, $2560\times 1440$ pixels at 60~FPS
is equivalent to $4741~\mu$s/Mpixel.)}

Finally, \chadded{we note that}
our method can \chadded{similarly} reduce the computational cost even with traditional texture representations.
Pharr et al.~\cite{Pharr2024} discussed how STF can be used to accelerate common expensive
material evaluation techniques requiring multiple material samples, such as triplanar mapping.
Our method can be used to similarly distribute sampling of different axis-aligned planes between lanes in a wave,
or, in general, even different material samples.

\section{Discussion and Limitations}
\label{sec_discussion}

Our results in Section~\ref{sec_results_bilinear} suggest that \WM gives the highest quality
independent of whether denoising is used,
though \WB does almost as well with a lower runtime cost. 
This makes \WB our general recommendation.
However, because \WM gives a more accurate estimate of which texels are required
for perfect filtering,
it is an attractive alternative when combined with larger filters (Section~\ref{sec_results_bicubic})
where the fallback method may be necessary less often
or with decompression methods that split decompression work across lanes (Section~\ref{sec_perf_expensive_decompressors}).

\chdeleted{
Our methods often produce noise-free filtered textures and, as a result, noise-free material properties.
Since both traditional and neural denoisers use material properties in guide buffers or for demodulation,
noise-free material buffers improve denoised results.
However, although our fallback methods give higher-quality images than the state of the art
when stochastic evaluation is required, their results may require denoising
and the consequences of filtering after shading may still be observed.
}

We have shown that our algorithms guarantee perfect filtering with small discrete texture filters 
\chreplaced{above}{after} modest magnification \chadded{factors} and also do so
with larger texture filters at higher magnification factors (Section~\ref{sec_bicubic_zoom}).
This is a natural outcome, since
more texels are needed for larger texture filters.
For perfect filtering at lower magnification rates,
all of our methods can easily be extended to generate more than one texel per lane.
We have verified experimentally that for a screen-\chreplaced{aligned}{space} quad with magnification
factor $1.0$, \chdeleted{then}the bilinear filter requires no more than 54 unique texels 
(\chadded{occuring }at\chadded{, e.g.,} $30^\circ$ rotation of the quad) for $8\times 4$
waves. 
Hence, $\leq 2$ texel \chreplaced{evaluations}{lookups} are sufficient
to \textit{always} succeed at returning a perfectly bilinearly-filtered value \chreplaced{when all pixels in a wave undergo magnification}{under
magnification}. Similar analysis for larger filters is left for
future work.

While we focused our work on the application of magnification filtering, we found that our
fallback methods apply to modest minification filters and trilinear filtering as well,
effectively performing filtering before shading.
We have decided not to include those results following the recommendations of Pharr et al.~\cite{Pharr2024},
as well as common real-time rendering practice of negative MIP biasing and relying on post-shading
spatiotemporal reconstruction filters~\cite{Karis:2014:High}.
Filtering before shading changes the appearance of rendered objects under minification,
reduces the perceived texture sharpness, and \chreplaced{can lead to aliasing}{leads to aliasing when using a highly nonlinear
shading function}. 

One limitation of our algorithms is a consequence of their reliance on obtaining
complete enumerable sets of texels to filter.
Pharr et al.~\cite{Pharr2024} proposed a second family of stochastic
texture filtering for continuous and potentially infinite filters, based on
filter importance sampling.
Determining the complete set of texels required for perfect filtering is inherently impossible for any filter
with an infinite spatial support, which makes our methods inapplicable.
To use our algorithms with such filters, truncation and windowing  
would be necessary, which would lead to approximation errors.
Furthermore, computing multiple filter weights is significantly more costly than sampling the filter PDF
in the case of commonly used filters, such as the Gaussian kernel.

\section{Conclusions and Future Work}
\label{sec_conclusions}
\chreplaced{Our}{We believe that our} family of techniques effectively addresses the shortcomings of 
stochastic texture filtering during magnification for commonly-used texture filters.
Above small magnification factors, \chreplaced{they}{our methods} achieve \chdeleted{filtering with} zero error while having a small computational cost,
\chdeleted{and require} generating no more than one texel per lane.

While we briefly investigated techniques
to increase parallelism for decompression when few unique texels are required 
(Section~\ref{sec_dct_decompression}),
further work in this direction may give even greater
performance improvements in such cases.
Extensions to other decompressors may give similar performance benefits.

In our approach, we rely on inter-wave communication to minimize performance overhead.
However, this limits the number of GPU threads collaboratively \chreplaced{producing}{fetching} texels to the wave size and limits
the applicability of our algorithm to small texture filters.
In future work, it might be worth extending our approach to a hybrid between wave communication
and using shared memory with larger compute shader launch groups.

\chdeleted{
Finally, our algorithmic contributions of the new \WB and \WM methods serve as conservative approximations of
the expensive \WL method; those approximations might be useful in other parallel computing
applications.
}

\section*{Acknowledgments}

Many thanks to
Tom{\'a}{\v s} Davidovi{\v c} for help with details of Falcor internals
and to
Johannes Deligiannis for help with integrating our techniques with the NTC SDK.
Magnus Andersson,
Rasmus Barringer,
Anders Lindqvist,
James Player, and
Robert Toth
all helped with code optimizations.
Thanks also to Aaron Lefohn and NVIDIA Research for supporting this work.

Thanks to \emph{PolyHaven} for the Aerial Rocks, Dirt, and Painted Concrete texture sets and
to \emph{ambientCG} for the Bricks and Rails texture sets.

\bibliographystyle{eg-alpha-doi} 
\bibliography{references}

\newcommand{\etalchar}[1]{$^{#1}$}
\begin{thebibliography}{\uppercase{WPAM25}}

\bibitem[ANA{\etalchar{*}}20]{Andersson2020}
\textsc{Andersson P., Nilsson J., Akenine{-}M{\"{o}}ller T., Oskarsson M.,
  {\AA}str{\"{o}}m K., Fairchild M.~D.}:
\newblock {\FLIP: A Difference Evaluator for Alternating Images}.
\newblock \emph{Proceedings of the ACM on Computer Graphics and Interactive
  Techniques 3}, 2 (2020), 15:1--23.

\bibitem[BWP{\etalchar{*}}20]{bitterli:2020:restir}
\textsc{Bitterli B., Wyman C., Pharr M., Shirley P., Lefohn A., Jarosz W.}:
\newblock {Spatiotemporal Reservoir Resampling for Real-Time Ray Tracing with
  Dynamic Direct Lighting}.
\newblock \emph{ACM Transactions on Graphics 39}, 4 (July 2020).
\newblock \href {https://doi.org/10/gg8xc7} {\path{doi:10/gg8xc7}}.

\bibitem[DBB{\etalchar{*}}25]{Dupuy2025}
\textsc{Dupuy J., Benyoub A., Belcour L., Merecki M., Chambon T.}:
\newblock {Intel Co-Presents Cooperative Vectors with Microsoft}.
\newblock Game Developers Conference, 2025.
\newblock [Online; accessed 2025-03-25].

\bibitem[FH24]{Fujieda2024}
\textsc{Fujieda S., Harada T.}:
\newblock {Neural Texture Block Compression}.
\newblock In \emph{Workshop on Material Appearance Modeling} (2024), Hardeberg
  J.~Y., Rushmeier H., (Eds.), The Eurographics Association.

\bibitem[Hoo11]{Hoobler2011}
\textsc{Hoobler N.}:
\newblock {High Performance Post-Processing}.
\newblock In \emph{Game Developers Conference} (2011).

\bibitem[Kar14]{Karis:2014:High}
\textsc{Karis B.}:
\newblock {High-Quality Temporal Supersampling}.
\newblock \emph{Advances in Real-Time Rendering in Games, SIGGRAPH Courses 1},
  10.1145 (2014), 2614028--2615455.

\bibitem[KCK{\etalchar{*}}22]{Kallweit2022}
\textsc{Kallweit S., Clarberg P., Kolb C., Davidovi{\v c} T., Yao K.-H., Foley
  T., He Y., Wu L., Chen L., Akenine{-}M{\"{o}}ller T., Wyman C., Crassin C.,
  Benty N.}:
\newblock The falcor rendering framework.
\newblock BSD-Licensed Github Repository, August 2022.

\bibitem[KKSS18]{kenzel2018high}
\textsc{Kenzel M., Kerbl B., Schmalstieg D., Steinberger M.}:
\newblock {A High-Performance Software Graphics Pipeline Architecture for the
  GPU}.
\newblock \emph{ACM Transactions on Graphics 37}, 4 (2018), 1--15.

\bibitem[KLM24]{Kim2024neuralvdb}
\textsc{Kim D., Lee M., Museth K.}:
\newblock {NeuralVDB: High-Resolution Sparse Volume Representation Using
  Hierarchical Neural Networks}.
\newblock \emph{ACM Transactions on Graphics 43}, 2 (2024), 20:1--21.

\bibitem[LSO07]{lefohn:2007:rmsm}
\textsc{Lefohn A.~E., Sengupta S., Owens J.~D.}:
\newblock {Resolution-Matched Shadow Maps}.
\newblock \emph{ACM Transactions on Graphics 26}, 4 (Oct. 2007), 20--37.

\bibitem[MHA{\etalchar{*}}24]{Mantiuk2024}
\textsc{Mantiuk R.~K., Hanji P., Ashraf M., Asano Y., Chapiro A.}:
\newblock {ColorVideoVDP: A Visual Difference Predictor for Image, Video and
  Display Distortions}.
\newblock \emph{ACM Transactions on Graphics 43}, 4 (2024), 129:1--20.

\bibitem[Mic21]{Microsoft2021}
\textsc{Microsoft}:
\newblock {HLSL Shader Model 6.0}.
\newblock
  \url{https://learn.microsoft.com/en-us/windows/win32/direct3dhlsl/hlsl-shader-model-6-0-features-for-direct3d-12},
  2021.
\newblock [Online; accessed 2024-09-11].

\bibitem[MML12]{mcguire:2012:scalable}
\textsc{McGuire M., Mara M., Luebke D.}:
\newblock {Scalable Ambient Obscurance}.
\newblock In \emph{High Performance Graphics} (2012), pp.~97--103.

\bibitem[NVI25]{NVIDIA2025:DLSS}
\textsc{NVIDIA}:
\newblock {DLSS 4: Transforming Real-Time Graphics with AI}.
\newblock \url{https://research.nvidia.com/labs/adlr/DLSS4/}, 2025.
\newblock Technical Report.

\bibitem[OBA12]{olsson:2012:clustered}
\textsc{Olsson O., Billeter M., Assarsson U.}:
\newblock {Clustered Deferred and Forward Shading}.
\newblock In \emph{High Performance Graphics} (2012), pp.~87--96.

\bibitem[Pen11]{penner:2011:shader}
\textsc{Penner E.}:
\newblock {Shader Amortization Using Pixel Quad Message Passing}.
\newblock In \emph{GPU Pro 2}. CRC Press, 2011, pp.~349--366.

\bibitem[PWSF24]{Pharr2024}
\textsc{Pharr M., Wronski B., Salvi M., Fajardo M.}:
\newblock {Filtering After Shading with Stochastic Texture Filtering}.
\newblock \emph{Proceedings of the ACM on Computer Graphics and Interactive
  Techniques 7}, 1 (2024), 14:1--20.

\bibitem[SHG09]{satish:2009:sort}
\textsc{Satish N., Harris M., Garland M.}:
\newblock {Designing Efficient Sorting Algorithms for Manycore {GPU}s}.
\newblock In \emph{IEEE International Symposium on Parallel \& Distributed
  Processing} (2009), pp.~1--10.

\bibitem[SHGO11]{sengupta:2011:efficient}
\textsc{Sengupta S., Harris M.~J., Garland M., Owens J.~D.}:
\newblock {Efficient Parallel Scan Algorithms for Many-Core {GPU}s}.
\newblock In \emph{Scientific Computing with Multicore and Accelerators},
  Jakub~Kurzak D. A.~B., Dongarra J., (Eds.). 2011, pp.~413--442.

\bibitem[VSW{\etalchar{*}}23]{Vaidyanathan2023}
\textsc{Vaidyanathan K., Salvi M., Wronski B., Akenine{-}M{\"{o}}ller T.,
  Ebelin P., Lefohn A.}:
\newblock {Random-Access Neural Compression of Material Textures}.
\newblock \emph{ACM Transactions on Graphics 42}, 4 (2023), 88:1--25.

\bibitem[WPAM25]{Wronski2025}
\textsc{Wronski B., Pharr M., Akenine-M\"oller T.}:
\newblock {Improved Stochastic Texture Filtering Through Sample Reuse}.
\newblock \emph{Proceedings of the ACM on Computer Graphics and Interactive
  Techniques 8}, 1 (2025).
\newblock \href {http://arxiv.org/abs/2504.05562} {\path{arXiv:2504.05562}}.

\bibitem[YLS20]{Yang:2020:Survey}
\textsc{Yang L., Liu S., Salvi M.}:
\newblock {A Survey of Temporal Antialiasing Techniques}.
\newblock \emph{Computer Graphics Forum 39}, 2 (2020), 607--621.

\end{thebibliography}

\SupplementaryMaterials
\vfill\eject
\pagebreak
\begin{center}
  \textbf{\large Supplemental Material:\\Collaborative Texture Filtering}
\end{center}

\noindent
This supplemental material includes:
\begin{itemize}
\item Detailed code showing example implementations of the \WB and \WM algorithms.
\item Details on the edge remapping technique used at silhouette edges where not all of the wave's lanes
are active.
\item More information about the scenes, textures, and metrics used for evaluation.
\item Additional evaluation and results.
\end{itemize}

\definecolor{codegreen}{rgb}{0,0.6,0}
\definecolor{codegray}{rgb}{0.5,0.5,0.5}
\definecolor{backcolour}{rgb}{0.95,0.95,0.92}

\lstdefinestyle{mystyle}{
       backgroundcolor=\color{backcolour},
       commentstyle=\color{codegreen},
       keywordstyle=\color{magenta},
       numberstyle=\tiny\color{codegray},
       stringstyle=\color{codepurple},
       basicstyle=\ttfamily\scriptsize,
       breakatwhitespace=false,
       breaklines=true,
       captionpos=b,
       keepspaces=true,
       numbers=left,
       numbersep=5pt,
       showspaces=false,
       showstringspaces=false,
       showtabs=false,
       tabsize=2
}

\lstset{style=mystyle}
 
\section{Code}
\label{supp_pseudocode}

For simplicity, we omit clamping of out-of-bounds texture coordinates in the code below
and we do not include an implementation of \WL, since it performs poorly and is not a realistic
alternative for use in real applications. (Recall that it was mostly
included because it computes the number of texels needed for perfect filtering exactly,
which was useful for evaluating the other approaches.)

\subsection{\WBc}
The following helper functions are used in the implementation of \WB for bilinear filtering, which follows.
\lstdefinestyle{mystyle}
{
	language = C,
	keywordstyle = [1]{\color{magenta}},
	keywordstyle = [2]{\color{violet}},
	morekeywords = [1]{float4, uint, uint2, int2, uint64_t4},
	morekeywords = [2]{WaveGetLaneIndex, WaveReadLaneAt, WaveActiveMin, WaveActiveMax, WaveActiveBitOr, GetDimensions, WaveActiveBallot, countbits, fallBackMethod, computeBilinearWeights, texture, bijectiveFunctionH, LanesLowerThanCountActive, CoordToLaneIdx, LaneIdxToCoord},
	basicstyle=\tiny\ttfamily,
	frame=single,
}
\begin{lstlisting}[language=C,  style=mystyle]
int2 LaneIdxToCoord(uint laneIdx,
					int2 waveUpperLeftIntCoords,
					uint bbWidth)
{
    uint laneY = laneIdx / bbWidth;
    uint laneX = laneIdx %
    return waveUpperLeftIntCoords + int2(laneX, laneY);
}

uint CoordToLaneIdx(int2 coord,
					int2 waveUpperLeftIntCoords,
					uint bbWidth)
{
    coord -= waveUpperLeftIntCoords;
    return coord.x + coord.y * bbWidth;
}

bool LanesLowerThanCountActive(uint count)
{
    uint activeLanesMask = WaveActiveBallot(true).x;
    uint desiredActiveMask = (count == 32) ? 0xFFFFFFFF :
        (1 << count) - 1;

    return (activeLanesBitMask & desiredActiveMask) == desiredActiveMask;
}
\end{lstlisting}
Inside the shader that accesses the texture, the following code implements \WB:
\lstdefinestyle{mystyle}
{
	language = C,
	keywordstyle = [1]{\color{magenta}},
	keywordstyle = [2]{\color{violet}},
	morekeywords = [1]{float4, float2, uint, uint2, int2, uint64_t4, bool},
	morekeywords = [2]{WaveGetLaneIndex, WaveReadLaneAt, WaveActiveMin, WaveActiveMax, WaveActiveBitOr, GetDimensions, WaveGetLaneCount, countbits, fallBackMethod, computeBilinearWeights, texture, bijectiveFunctionH, LanesLowerThanCountActive, CoordToLaneIdx, LaneIdxToCoord},
	basicstyle=\tiny\ttfamily,
	frame=single,
}
\begin{lstlisting}[language=C,  style=mystyle]
uint2 txDim;
texture.GetDimensions(txDim.x, txDim.y);
float2 floatCoords = uv * txDim - float2(0.5f);
int2 upperLeftIntCoords = int2(floor(floatCoords));
int2 lowerRightIntCoords = upperLeftIntCoords + int2(1, 1);
float2 stCoords = floatCoords - upperLeftIntCoords;

// Compute bounding box of texel integer coordinates
int2 waveUpperLeftCoords = WaveActiveMin(upperLeftIntCoords);
int2 waveLowerRightCoords = WaveActiveMax(lowerRightIntCoords);
int2 bbSize = waveLowerRightCoords - waveUpperLeftCoords + 1;

int activeTexelsNeeded = bbSize.x * bbSize.y;
bool requiredLanesActive = LanesLowerThanCountActive(activeTexelsNeeded);
	
if (activeTexelsNeeded > 32 || !requiredLanesActive)
	return fallBackMethod();

uint curLaneIdx = WaveGetLaneIndex();
float4 texelValue = float4(0.0f);

if (curLaneIdx <= activeTexelsNeeded) {
	uint2 texCoords = LaneIdxToCoord(curLaneIdx,
                                   waveUpperLeftCoords, bbSize.x);
	texelValue = texture[texCoords];
}

float4 bilinWeights = computeBilinearWeights(stCoords);

float4 filteredColor = float4(0.0f);
[unroll]
for (int i = 0; i < 4; i++) {
	int2 texelCoords = int2(upperLeftIntCoords.x + (i %
                          upperLeftIntCoords.y + (i / 2));
	uint laneIdx = CoordToLaneIdx(texelCoords, waveUpperLeftCoords, bbSize.x);
	filteredColor += WaveReadLaneAt(texelValue, laneIdx) * bilinWeights[i];
}
return filteredColor;
\end{lstlisting}
The first six lines of code compute various coordinates from 
the texture dimensions and $(u,v)$-coordinates.
Among these are the upper left coordinates
of the $2\times 2$ filter footprint needed for bilinear filtering
and $(s,t)$-coordinates which are local coordinates inside
the $2\times 2$ region and are in $[0,1]$.
Next, we use \texttt{WaveActiveMin()} and \texttt{WaveActiveMax()}
to compute an
axis-aligned bounding box over the texel coordinates that are
needed for the entire wave.

Next, we check whether the number of needed texels is less than
or equal to the number of active lanes in the wave and if all of the lanes are active.
For each texel-producing lane, we then use a simple linearly-ordered mapping 
\texttt{LaneIdxToCoord()} to compute its texel coordinates.
Producing a texel in this example is done here via a regular texel fetch using
\texttt{texture[]}.
Finally, we iterate over the texture filtering footprint, gathering and accumulating the necessary 
texels using the \texttt{CoordToLaneIdx()} and \texttt{WaveReadLaneAt()} functions.
We note that the same algorithm works for any texture filter:
the only changes required are the calculations of \texttt{upperLeftIntCoords}, \texttt{lowerRightIntCoords},
\texttt{computeBilinearWeights}, and the final unrolled loop extent.

\subsection{\WMc}
Code for \WM with a $16 \times 16$ mask is shown below.
\lstdefinestyle{mystyle}
{
	language = C,
	keywordstyle = [1]{\color{magenta}},
	keywordstyle = [2]{\color{violet}},
	morekeywords = [1]{float2, float4, uint, uint2, int2, uint64_t4},
	morekeywords = [2]{WaveGetLaneIndex, WaveReadLaneAt, WaveActiveMin, WaveActiveMax, WaveActiveBitOr,
		WaveActiveBallot, GetDimensions, countbits, fallBackMethod, computeBilinearWeights, texture, bijectiveFunctionH,inverseBijectiveFunctionH, set2x2Bits},
	basicstyle=\tiny\ttfamily,
	frame=single,
}
\begin{lstlisting}[language=C,  style=mystyle]
uint2 txDim;
texture.GetDimensions(txDim.x, txDim.y);
floatCoords = uv * txDim - float2(0.5f);
int2 upperLeftIntCoords = int2(floor(floatCoords));
int2 lowerRightIntCoords = upperLeftIntCoords + int2(1, 1);
float2 stCoords = floatCoords - upperLeftIntCoords;

// Compute bounding box of texel integer coordinates
int2 waveUpperLeftCoords = WaveActiveMin(upperLeftIntCoords);
int2 waveLowerRightCoords = WaveActiveMax(lowerRightIntCoords);
int2 bbSize = waveLowerRightCoords - waveUpperLeftCoords + 1;
uint2 deltaCoords = upperLeftIntCoords - waveUpperLeftIntCoords;

if (bbSize.x > 16 || bbSize.y > 16)
	return fallBackMethod();
	
// Set the 2x2 bits corresponding to the 2x2 texels needed.
uint64_t4 mask = uint64_t4(0);
set2x2Bits(upperLeftIntCoords - waveUpperLeftIntCoords, mask);

// Compute the ORed mask across the entire wave.
uint64_t4 waveMask;
waveMask.x = WaveActiveBitOr(mask.x);
waveMask.y = WaveActiveBitOr(mask.y);
waveMask.z = WaveActiveBitOr(mask.z);
waveMask.w = WaveActiveBitOr(mask.w);

uint curLaneIdx = WaveGetLaneIndex();
uint activeTexelsNeeded = countbits(waveMask);
uint activeLanesMask = WaveActiveBallot(true).x; // # active lanes

if (activeTexelsNeeded > countbits(activeLanesBitMask))
	return fallBackMethod();

float4 curPixelTexelValue;
int2 sampledTexelIntCoords;
float4 color = float4(0.0f);
float4 bilinWeights = computeBilinearWeights(stCoords);

// Does this lane needs to produce a texel?
if (curLaneIdx <= activeTexelsNeeded) {
	uint2 deltaTexCoords = bijectiveFunctionH(curLaneIdx, waveMask);
	sampledTexelIntCoords = waveUpperLeftIntCoords + deltaTexCoords;
	curPixelTexelValue = texture[sampledTexelIntCoords];
}

// Compute 1D local texture coordinate:
uint t = (deltaCoords.y << 4) + deltaCoords.x;
// Read the 2x2 texels needed for this pixel and weight together.
uint i0 = inverseBijectiveFunctionH(t, waveMask);
color += bilinWeights.x * WaveReadLaneAt(curPixelTexelValue, i0);
uint i1 = inverseBijectiveFunctionH(t+1, waveMask);
color += bilinWeights.y * WaveReadLaneAt(curPixelTexelValue, i1);
uint i2 = inverseBijectiveFunctionH(t+16, waveMask);
color += bilinWeights.z * WaveReadLaneAt(curPixelTexelValue, i2);
uint i3 = inverseBijectiveFunctionH(t+16+1, waveMask);
color += bilinWeights.w * WaveReadLaneAt(curPixelTexelValue, i3);
return color;
\end{lstlisting}
The first six lines of code are the same as for \WB.
The mask is stored in an \texttt{uint64\_t4}, which has $16\cdot 16=256$ bits,
and so lines 14--15 call the fallback method if either of the bounding box
dimensions is larger than 16.
Lines 18--26 first set the $2\times 2$ bits of the texels needed
by the current lane in \texttt{mask} using a function
\texttt{set2x2Bits()}; then it is OR of all of these
over the entire wave using \texttt{WaveActiveBitOr()}
that gives the full wave mask, \texttt{waveMask}.

Lines 28--33 compute how many texels are needed by counting 
the set bits in \texttt{waveMask} and use \texttt{WaveActiveBallot(true)}
to find a bitmask of the active lanes in the wave. As with \WB,
the fallback method is called if the number of texels needed 
is more than the number of active lanes in the wave.
Past this point, we know there are a sufficient number of active
lanes and that \WM will succeed at getting all
$2\times 2$ texels needed by each active lane in the wave
to perform perfect bilinear filtering.

In lines 41--45, the lanes with lane number less than the
number of needed texels produce a texel, which in this
example simply does a texture lookup using the GPU's texture unit.
The bijective function $h(i,B)$ (Figure~\ref{fig_bijective_function_h})
is used to compute local
texture coordinates, which are added to the wave's upper left
coordinates and the lookup is performed at the end.
Line 48 computes the local one-dimensional texture coordinate
of the upper left texel for the current lane.
Since we know that the current lane wants the $2\times 2$
texels starting from \texttt{upperLeftIntCoords},
we can transform that to $2\times 2$ one-dimensional texture
coordinates as $\{t, t+1, t+16, t+16+1\}$,
where $t+1$ identifies the texel to the right of $t$,
and $t+16$ the texel below $t$, since the mask
is $16\times 16$ bits.
These coordinates are then fed to the inverse of
the bijective function, i.e., $h^{-1}(t,B)$,
and then the texel is \chreplaced{produced}{looked up} from that
lane and weighted,
with perfect bilinear filtering as a result.

Similar to \WBc, we note that the algorithm above
works for any texture filter, only requiring changes to
the calculations of \texttt{upperLeftIntCoords}, 
\texttt{lowerRightIntCoords},
\texttt{computeBilinearWeights}, \texttt{set2x2Bits},
and the final unrolled loop.

\section{Edge Remapping}
\label{sec_remapping}
\begin{figure}
	\setlength{\tabcolsep}{0.75pt}
	\centering
	\begin{tabular}{ccc}		
		\includegraphics[width=0.32\columnwidth]{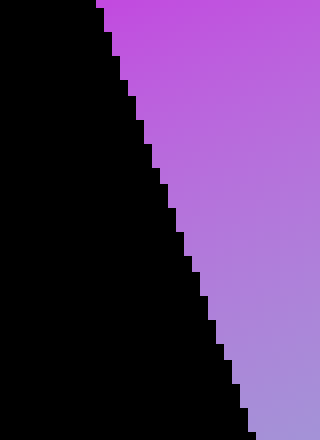}&
		\includegraphics[width=0.32\columnwidth]{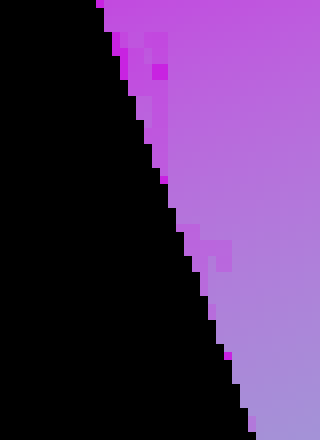}&
		\includegraphics[width=0.32\columnwidth]{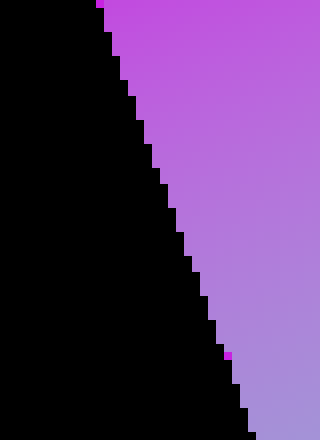}
		\\
		bilinear ground truth & \WM w/o  & \WM with
		\\
		& edge remapping & edge remapping
	\end{tabular}
	\vspace{0.75em}
	\caption{Left: bilinear ground truth filtering near a triangle
	  edge. Middle: if we require the first $n$ lanes to be active if
          $n$ texels are needed, then the fallback is often used and
          results may be noisy.
	Right: with the remapping technique described in Section~\ref{sec_remapping}, we
        only need $n$ active lanes across the wave and this problem
	is largely eliminated.}
	\label{fig_remapping}
\end{figure}

Close to a triangle edge against the background,
a wave will typically not have all of its lanes active.
So far, we have assumed that at least the
first $n$ lanes are active when $n$ texels are needed
and so the mappings from lanes to texels in Box and \WM
may end up with inactive lanes being assigned to generate texels
but then not actually doing so.
In such cases, we may use the fallback method introduced in
Section~\ref{sec_fallback_method} of the main paper,
though that can give artifacts as shown in the middle
in Figure~\ref{fig_remapping}.

Alternatively, we can map lanes to texels more carefully, accounting for
inactive lanes.
Assume for now that the size of a wave is 8: 
if all lanes are active, then we have 
the active lane mask \texttt{11111111}.
In that case \WM, for example,
can simply use its bijective function, $t=h(i,B)$,
to map a lane index into a texel coordinate.
This fails if the active lane mask is,
for example, \texttt{11101010}. 
Still, if only four lanes are needed, then we can see that
since the number of bits in the mask is $\geq 4$, it
should be possible to run the algorithm anyway.

\begin{figure*}[t]
	\centering
	\newcommand{\mywidth}{23.5mm}
	\setlength{\tabcolsep}{0.75pt}
	\renewcommand{\arraystretch}{0.5}
	\begin{tabular}{cccccc}
		& & & & \textsc{Painted-}
		\\
		&
		\textsc{AerialRocks} & \textsc{Bricks} & \textsc{Dirt} & %
		\textsc{Concrete} & \textsc{Rails}
		\\
		\rotatebox{90}{Diffuse} &
		\includegraphics[width=\mywidth]{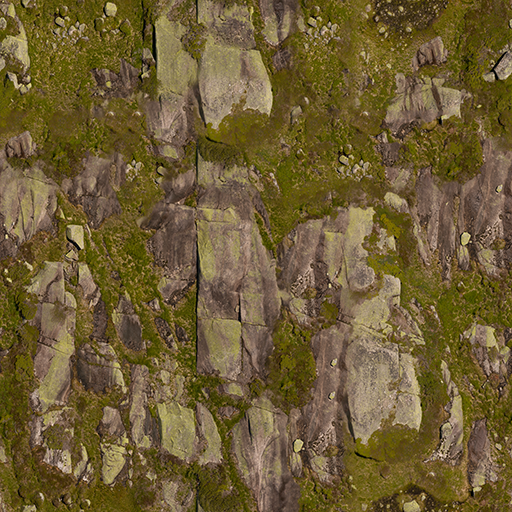} &
		\includegraphics[width=\mywidth]{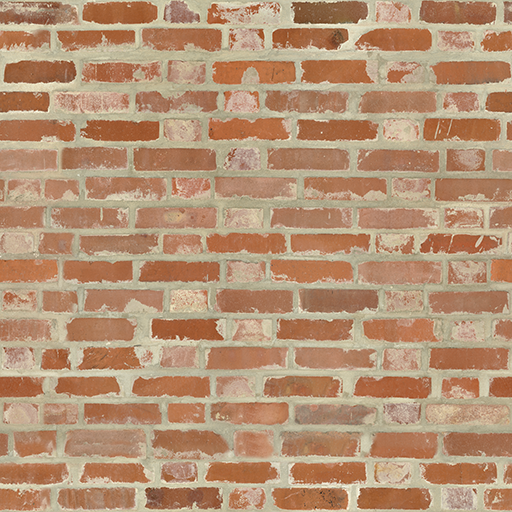} &
		\includegraphics[width=\mywidth]{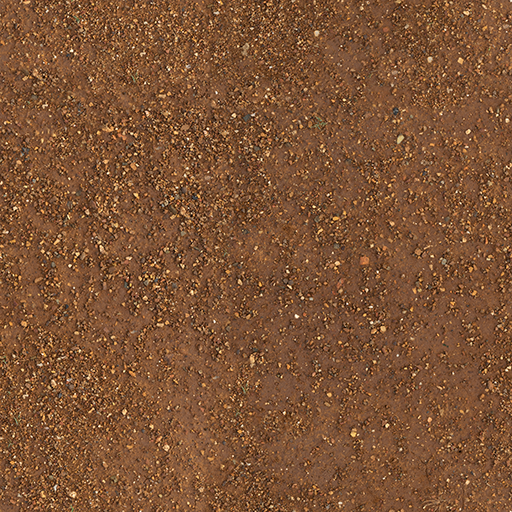} &
		\includegraphics[width=\mywidth]{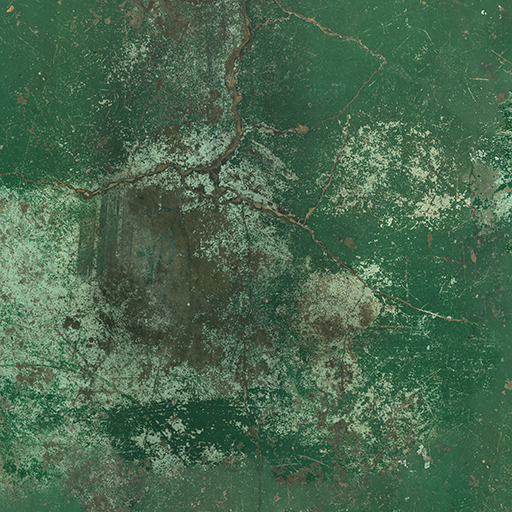} &
		\includegraphics[width=\mywidth]{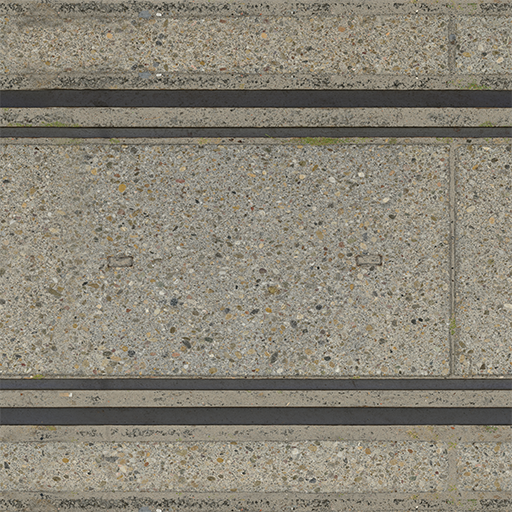}
		\\
		\rotatebox{90}{Normal map} &
		\includegraphics[width=\mywidth]{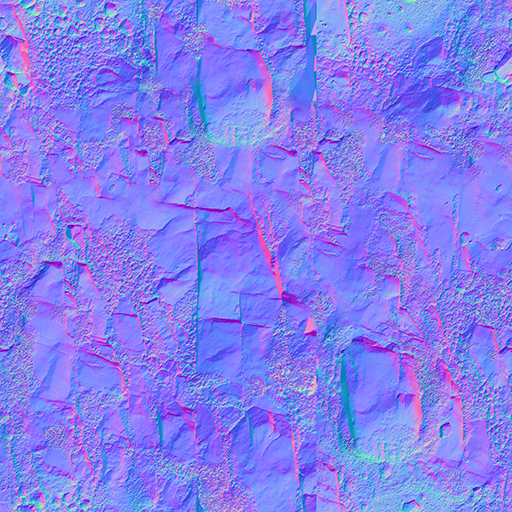} &
		\includegraphics[width=\mywidth]{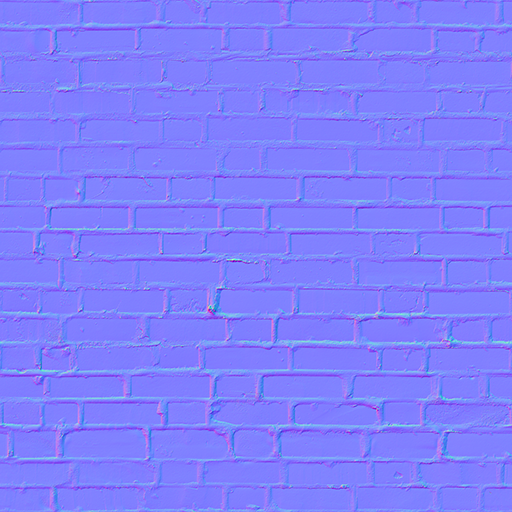} &
		\includegraphics[width=\mywidth]{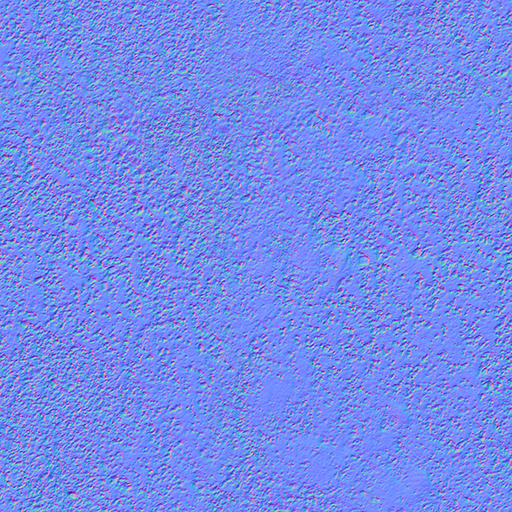} &
		\includegraphics[width=\mywidth]{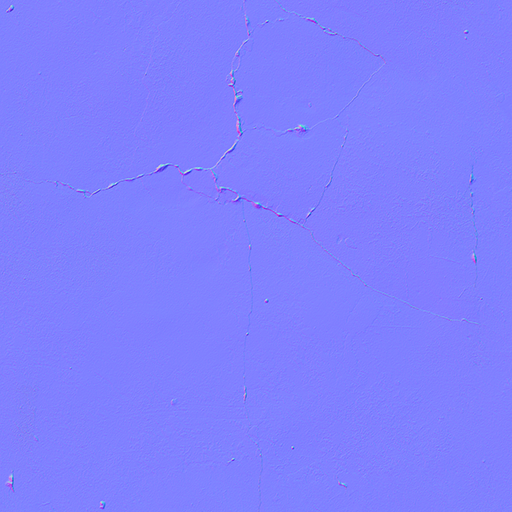} &
		\includegraphics[width=\mywidth]{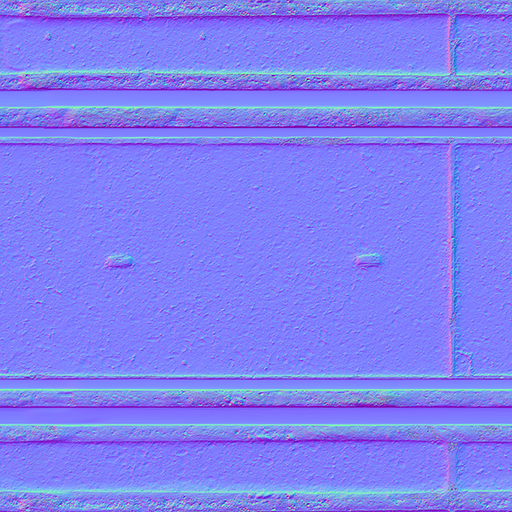}
	\end{tabular}
	\vspace*{0.75em}
	\caption{The textures used in our image quality measurements.}
	\label{fig_textures}
\end{figure*}

If we enumerate the needed texels from $0$ to $n-1$ with a number $i$,
then we need a way to map that number into a lane number
in the active lane mask where there is a \texttt{1}.
This can be done with the same bijective
function as is used in \WM.
We simply use $h()$ once again, i.e., we
remap so the texel number is remapped to lane index
as $h(i, A)$, where $A$ is the active lane mask.
This technique can also be applied to \WBc.

On the right side of Figure~\ref{fig_remapping},
we present the result of our edge remapping, which substantially improves the image quality.
However, it is still sometimes necessary to use the fallback method.
This can happen, for example, when two texels are needed (e.g., due to
clamping), but there is only a single active lane in the wave.

\section{Additional Evaluation Details and Results}
\label{supp_results}

In this section, we provide specifics
about the scenes and textures used in the paper
as well as the image quality metrics we used.
We also present additional quality results.

\subsection{Scenes and Textures}
\label{sec:scenes}

We used two primary scenes and six textures for evaluation.

The first test scene consisted of a single quad, and a camera positioned directly above the
quad's center, viewing it head on. The degrees of freedom used for this
scene was the camera's distance to the quad, which set the effective magnification factor,
the quad's rotation around the axis corresponding to the camera's viewing direction,
as well as the texture used for the quad.
The quad was shaded with a simple
shading algorithm that only considered diffuse and normal textures, with lighting defined by
a single light vector. This scene was used for Figures~\ref{fig_lookups}, \ref{fig:quality_zoom},
\ref{fig:supp_quality_zoom}, \ref{fig:rotation_max_error}, \ref{fig_fallback_evaluation},
\ref{fig_spp}, and \ref{fig:bicubic_lookups},
as well as Table~\ref{tab:rotating_errors}.

To analyze the behavior of our algorithms with partially active waves and texture discontinuities,
our second scene included \chreplaced{$4\times 4$}{four} tessellated spheres in addition to the quad.
The camera started some distance away from the quad
so that a small amount of minification occurred,
and then moved toward the spheres and back. The spheres
and the quad used the same texture, though the version used on the spheres
was a low-resolution version of the texture, so that we would have magnification
also on the spheres.
Each mesh in this scene was shaded using the same shading algorithm as the
quad in the first scene.
In total, the camera animation was \chreplaced{6}{14} seconds long and rendered
at 60 frames per second. It is shown in our supplemental video.
This scene was used for Figures~\ref{fig:paretoDLSS} and \ref{fig:pareto},
as well as Table~\ref{tab:results_table_combined}. A version
similar to this scene, but with different camera animations
and number of spheres, was used for Figures~\ref{fig_teaser} and
\ref{fig:bicubic}.

The resolution of the images we rendered was $1600\times 1600$,
and the field of view of the camera was 45 degrees.

The five textures used throughout the results section of this work were
the diffuse and normal textures for the
\textsc{AerialRocks}, \textsc{Bricks}, \textsc{Dirt}, %
\textsc{PaintedConcrete}, and \textsc{Rails} sets.
This set of textures shows variety
in both diffuse and normal content, including both high- and low-frequency information.
Each texture image has a resolution
of $4096\times 4096$ and we present them in Figure~\ref{fig_textures}.

\subsection{Metric Specifics}
Our results contain PSNR values as well as output from the ColorVideoVDP~\cite{Mantiuk2024}
and \FLIP metrics~\cite{Andersson2020}.

For single images, PSNR was computed as usual by
\begin{equation}
	\label{eq:psnr_single}
	\textrm{PSNR}(\mathbf{G}, \mathbf{T})=-10\log{\left(\textrm{MSE}(\mathbf{G}, \mathbf{T})\right)},
\end{equation}
where $\mathbf{G}$ is the ground-truth image, $\mathbf{T}$ is the test image, and MSE is
mean-squared error. Both ground-truth
and test images were in linear sRGB space. In cases where we have $N$ image sequences, each
containing $M$ frames, we take the average of the individual frame pairs' MSEs before computing PSNR.
That is, if $\mathbf{\hat{G}}^i = \{\mathbf{G}^i_j\}_{j=0}^{M-1}$ and
$\mathbf{\hat{T}}^i = \{\mathbf{T}^i_j\}_{j=0}^{M-1}$ are the sets of $M$ ground-truth and test
frames for sequence $i$, $i\in\{0, 1, \dots, N-1\}$, we compute PSNR as
\begin{equation}
	\label{eq:psnr_multiple}
	\begin{aligned}
	&\textrm{PSNR}\left(\{\mathbf{\hat{G}}^i\}_{i=0}^{N-1}, \{\mathbf{\hat{T}}^i\}_{i=0}^{N-1}\right) \\
	&=	-10\log{\left(\frac{1}{NM}\sum_{i=0}^{N-1}\sum_{j=0}^{M-1}\textrm{MSE}(\mathbf{G}^i_j, \mathbf{T}^i_j)\right)}.
	\end{aligned}
\end{equation}
Notice that the computation in Equation~\ref{eq:psnr_multiple} gives the same result as that in
Equation~\ref{eq:psnr_single} when only a single frame pair is considered.

The ColorVideoVDP model requires information about the assumed observer's viewing conditions and display.
We use the default settings, and set the assumed number of frames displayed per second to 60 for all our image
sequences. The reproducibility information provided when running the metric was
``ColorVideoVDP v0.4.2, 75.4 [pix/deg], Lpeak=200, Lblack=0.2,
Lrefl=0.3979 [cd/m\textasciicircum 2], (standard\_4k).''
ColorVideoVDP's output is in Just-Objectionable Differences (JOD) units, scaled to have a maximum
value of 10. Only if the reference and test images/sequences are visually indistinguishable
under the assumed viewing conditions is the JOD value of the test equal to 10.
In cases where we compute one JOD over a set of image sequences, that JOD is the
average of the JODs for each individual sequence. Using the notation introduced above,
we have
\begin{equation}
	\label{eq:cvvdp_multiple}
	\begin{aligned}
	&\textrm{ColorVideoVDP}\left(\{\mathbf{\hat{G}}^i\}_{i=0}^{N-1}, \{\mathbf{\hat{T}}^i\}_{i=0}^{N-1}\right)\\
	&= \frac{1}{N} \sum_{i=0}^{N-1} \textrm{ColorVideoVDP}\left(\mathbf{\hat{G}}^i, \mathbf{\hat{T}}^i \right).
	\end{aligned}
\end{equation}
Note that ColorVideoVDP is also able to compare single images. We use this
capability for the results in Figure~\ref{fig:quality_zoom}.

Like ColorVideoVDP, \FLIP requires information about the assumed observer's viewing conditions.
In particular, it takes the observer's distance to the display as well as the display's width (in meters and
in pixels). To match the assumptions for ColorVideoVDP and \FLIP, we use the information
provided by ColorVideoVDP for its \texttt{standard\_4k} display, namely a distance to display
of 0.7472 meters and a display width of 0.664 meters and 3840 pixels. As \FLIP acts
on single images, and not image sequences, the \FLIP error we present for multiple image sequences
is the average of each frame pairs' \FLIP error, i.e.,
\begin{equation}
	\label{eq:flip_multiple}
	\begin{aligned}
		&\textrm{\FLIP}\left(\{\mathbf{\hat{G}}^i\}_{i=0}^{N-1}, \{\mathbf{\hat{T}}^i\}_{i=0}^{N-1}\right) \\
		&=	\frac{1}{NM}\sum_{i=0}^{N-1}\sum_{j=0}^{M-1}\textrm{\FLIP}(\mathbf{G}^i_j, \mathbf{T}^i_j).
	\end{aligned}
\end{equation}

\subsection{Quality Comparison Without Denoising}
\begin{figure}[t]
	\centering
	\includegraphics[width=\linewidth]{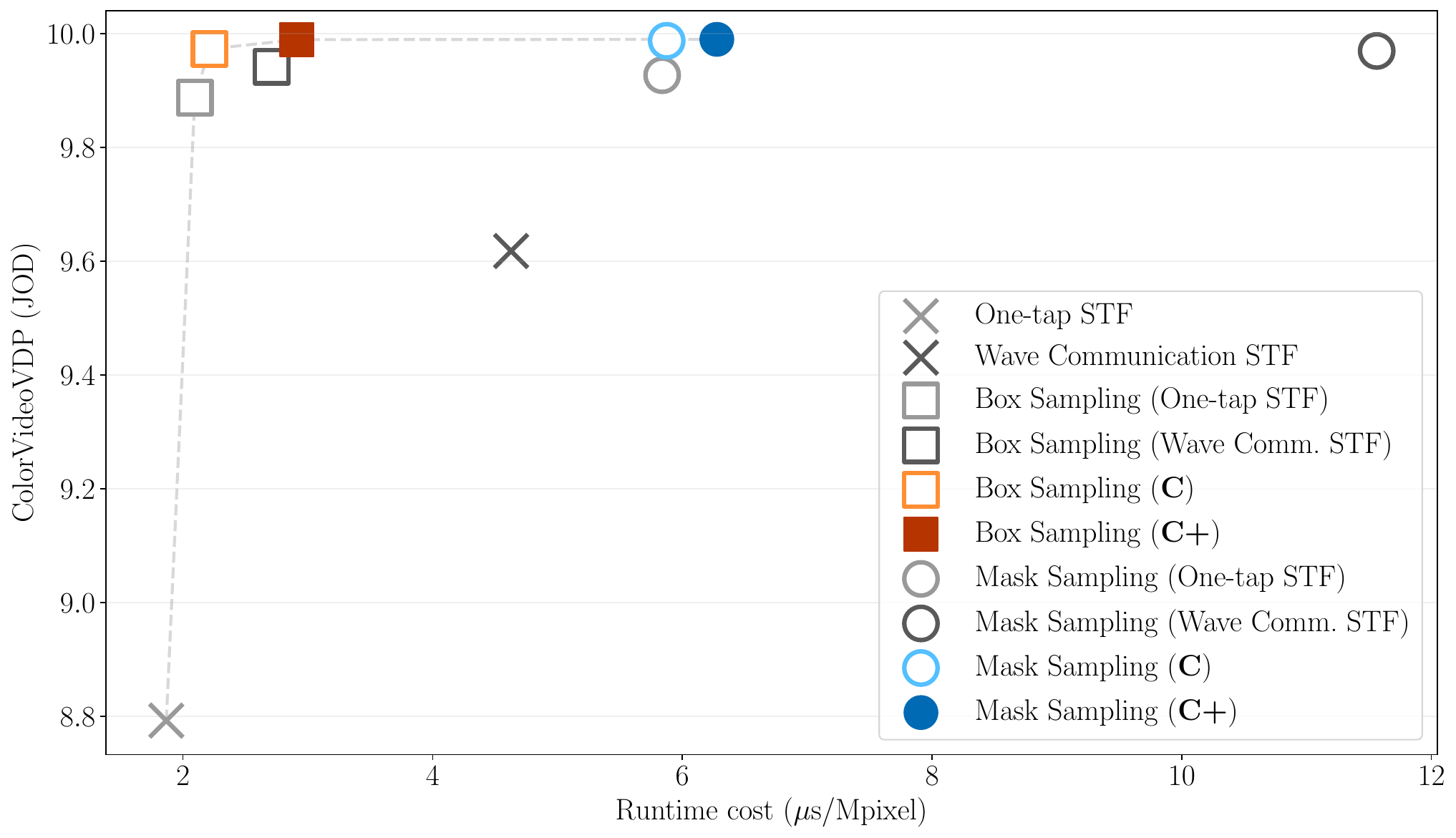}
	\vspace{0.75em}
	\caption{Pareto frontier (dashed line) indicating the most quality/performance-efficient
		algorithms when rendering without denoising. Runtime cost was measured
		on an NVIDIA RTX 5090. Our new fallback algorithms are marked with \textbf{C} and \textbf{C+}.
		The PSNR range for this plot was \chhighlight{29--59}\chdeleted{27--49}~dB.
		}
	\label{fig:pareto}
\end{figure}
\begin{figure}[t]
	\centering
	\includegraphics[width=\linewidth]{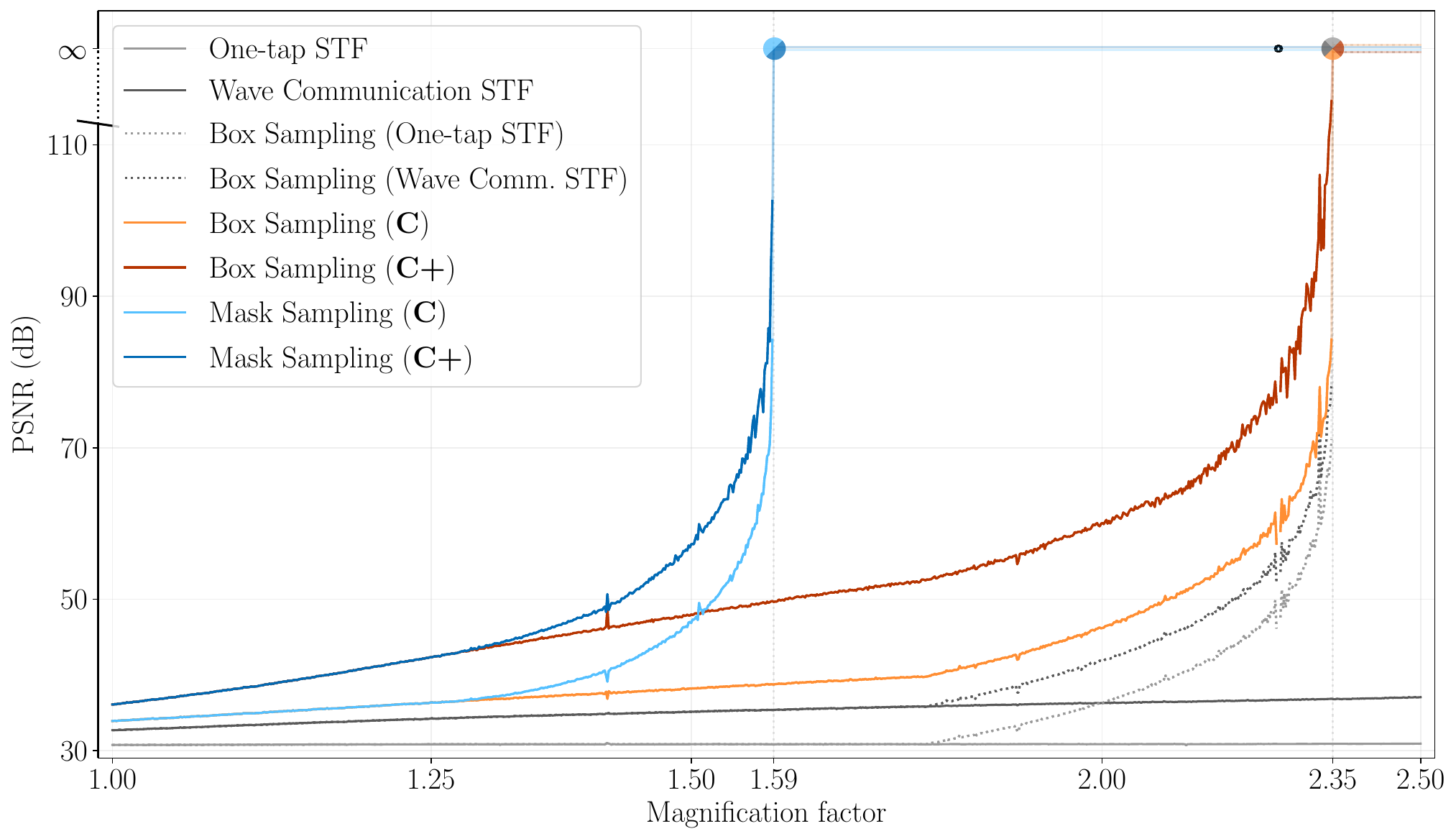}
	\vspace{0.75em}
	\caption{Quality as a function of magnification, without denoising.
		As indicated by Figure~\ref{fig_lookups},
		perfect bilinear filtering is achieved above magnification factor $2.35$
		for our methods.
		Colored discs mark that an algorithm has achieved perfect bilinear filtering at the corresponding magnification factor, resulting in
		infinite PSNR. At that and higher magnification factors, perfect bilinear filtering occurs for that algorithm.
		The small black circle marks a case when perfect bilinear filtering is achieved at a magnification lower than the limits
		indicated by the disks. This situation can arise for certain magnification and rotation combinations.
		}
	\label{fig:supp_quality_zoom}
\end{figure}

\begin{figure*}
	\centering
	\begin{subfigure}{\textwidth}
		\includegraphics[width=\textwidth]{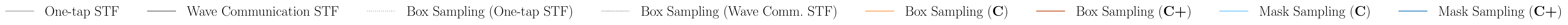}
	\end{subfigure}\\
	\begin{subfigure}{0.32\textwidth}
		\includegraphics[width=\textwidth]{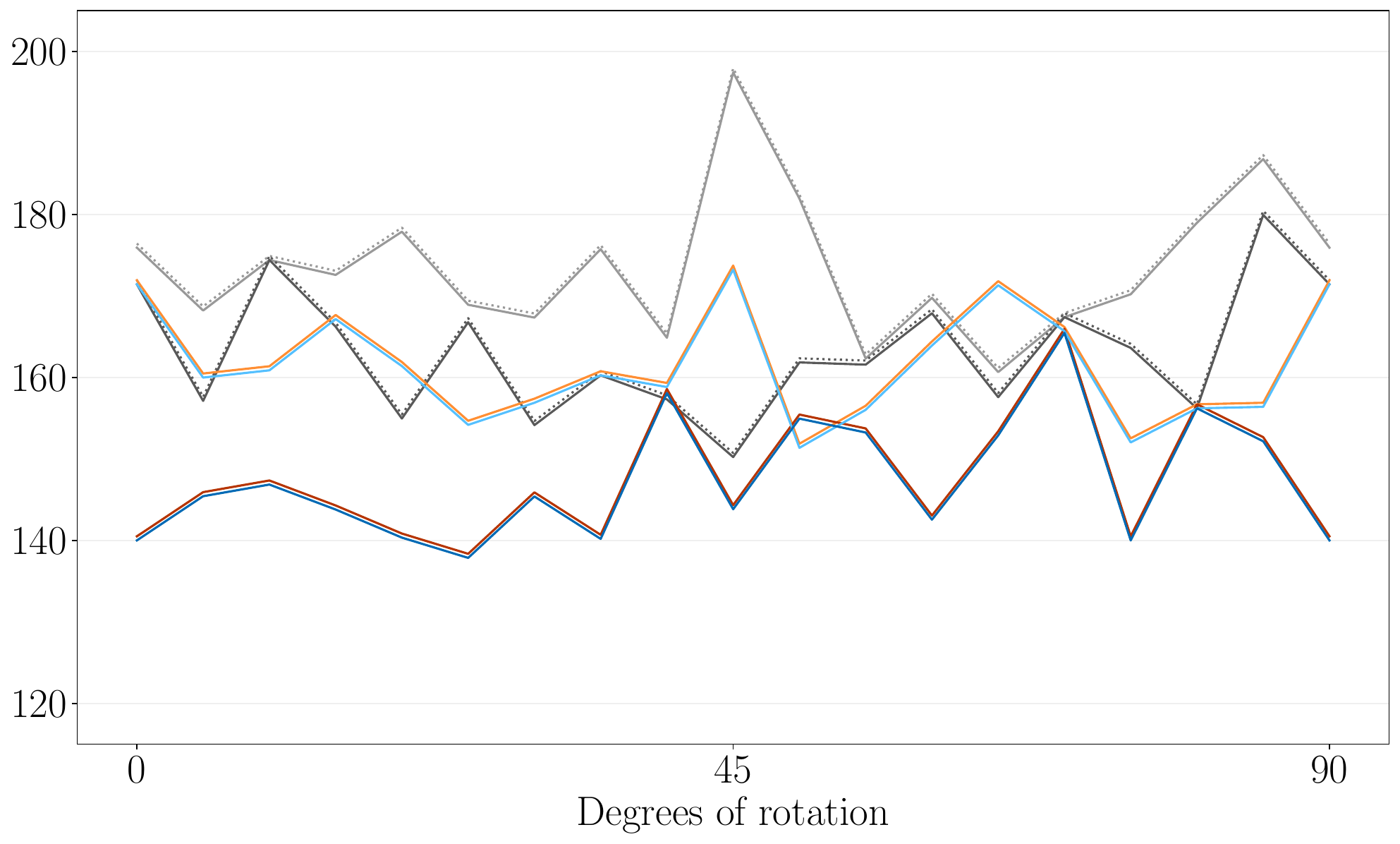}
		\caption{Magnification factor: 1.15}
		\label{fig:rotation_max_error_115}
	\end{subfigure}
	\begin{subfigure}{0.32\textwidth}
		\includegraphics[width=\textwidth]{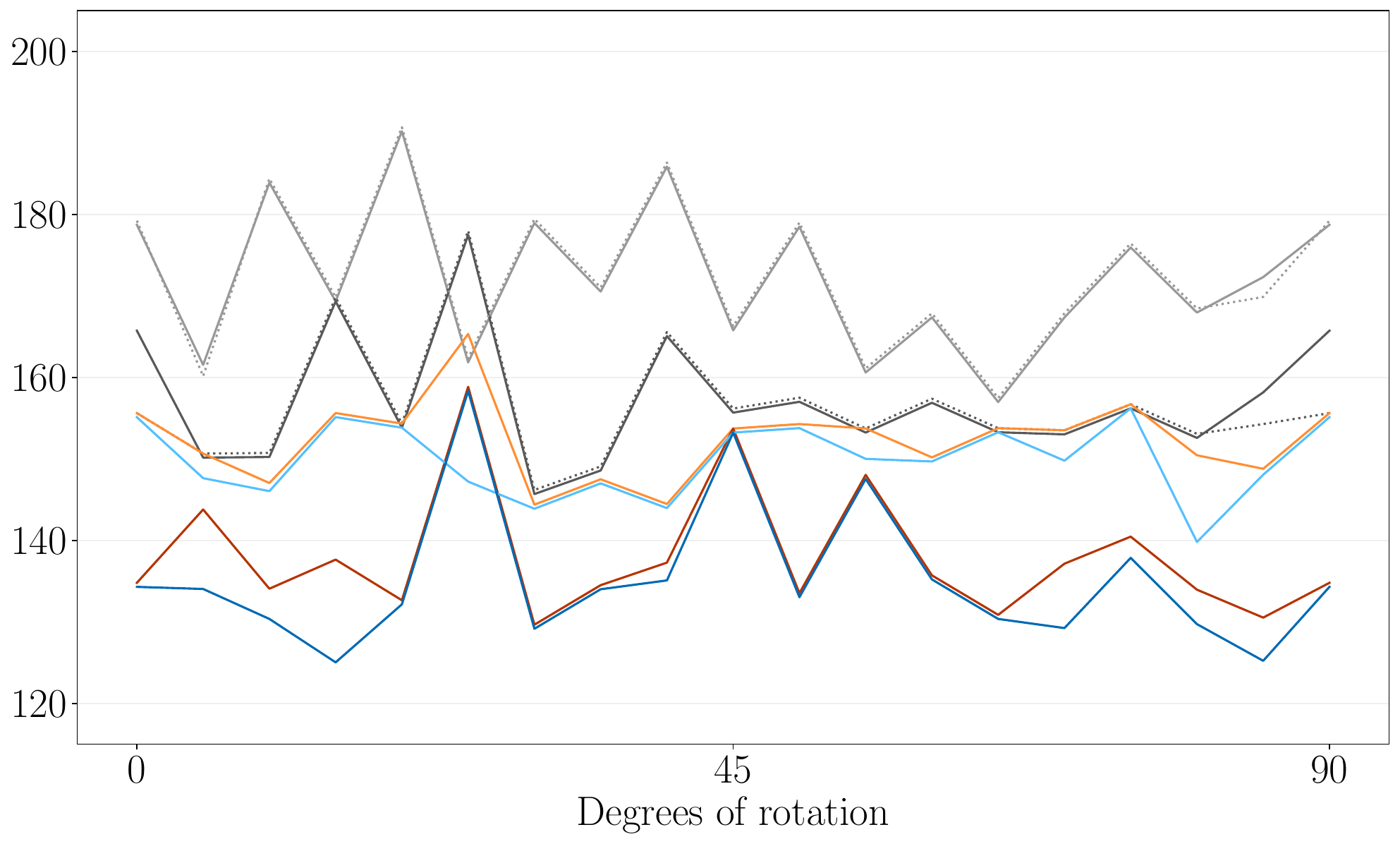}
		\caption{Magnification factor: 1.35}
		\label{fig:rotation_max_error_135}
	\end{subfigure}
	\begin{subfigure}{0.32\textwidth}
		\includegraphics[width=\textwidth]{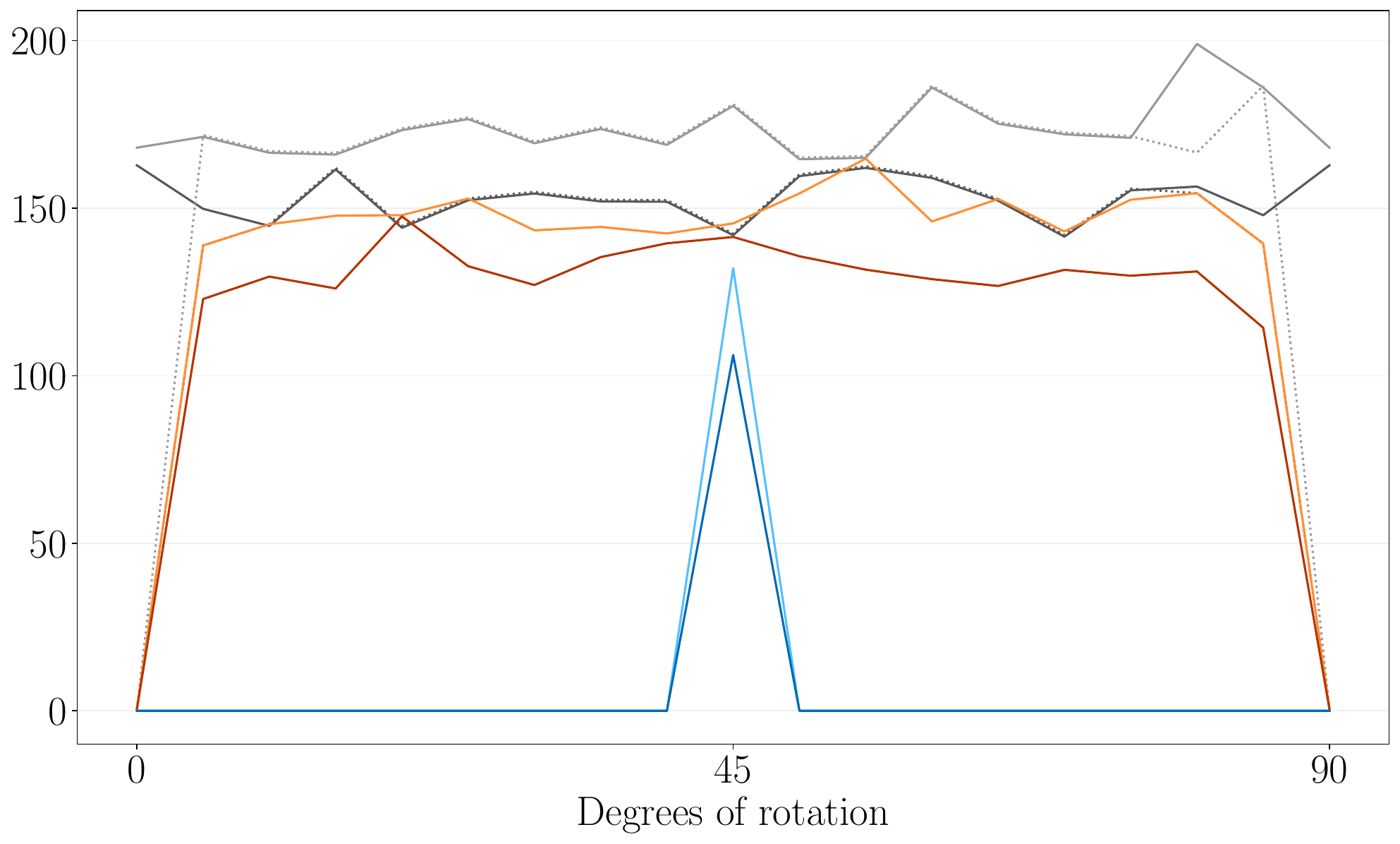}
		\caption{Magnification factor: 1.55}
		\label{fig:rotation_max_error_155}
	\end{subfigure}
	\vspace{0.75em}
	\caption{Maximum errors (scaled to $[0, 255]$) produced by the compared algorithms for rotations by $r\in[0,90]$ degrees,
		given three different levels of magnification. The legend is included above the plots.
		Note that the $y$-axis for the rightmost plot differs from that in the other two.
		The errors from Box and \WM in Figure~\ref{fig:rotation_max_error_115} are the same, but
		the curves for \WB are offset by 0.5 for visibility. As implied
		by Figure~\ref{fig_lookups}, at magnification factors larger than
		1.59, the maximum error of \WM is zero for all rotations, while \WB requires
		more magnification (2.35 or higher) to produce perfect bilinear filtering.
	}
	\label{fig:rotation_max_error}
	
	\vspace{1em}  %
	
	\definecolor{lightgray}{rgb}{0.9,0.9,0.9}
\captionof{table}{Average \emph{maximum} errors (scaled to $[0, 255]$) across the rotations used in Figure~\ref{fig:rotation_max_error}, for each of the different magnification factors used in that figure. \STF is the algorithm by Pharr et al.~\cite{Pharr2024}, Wave Comm.\ is short for \WC and is the algorithm by Wronski et al.~\cite{Wronski2025}, Box is short for \WBc, and Mask is short for \WMc.}
\label{tab:rotating_errors}
\vspace{0.75em}
\begin{tabular}{l|cccccccc}
\toprule
Zoom & \makecell{One-tap\\STF} & \makecell{Wave\\Comm.} & \makecell{Box\\(One-tap STF)} & \makecell{Box\\(Wave Comm.)} & \makecell{Box\\(\textbf{C})} & \makecell{Box\\(\textbf{C+})} & \makecell{Mask\\(\textbf{C})} & \makecell{Mask\\(\textbf{C+})} \\
\midrule
1.15 & {\color{rgb,1:red,0.6;green,0.6;blue,0.6}174} & {\color{rgb,1:red,0.35;green,0.35;blue,0.35}163} & {\color{rgb,1:red,0.6;green,0.6;blue,0.6}174} & {\color{rgb,1:red,0.35;green,0.35;blue,0.35}163} & {\color{rgb,1:red,1.0;green,0.5583632666810756;blue,0.19818011428336807}161} & {\color{rgb,1:red,0.7127015300411498;green,0.204225299979703;blue,0.0}\textbf{147}} & {\color{rgb,1:red,0.3277250527283637;green,0.7533866722110684;blue,1.0}161} & {\color{rgb,1:red,0.0;green,0.41394354779309744;blue,0.7090815962463488}\textbf{147}} \\
\arrayrulecolor{lightgray}\midrule\arrayrulecolor{black}
1.35 & {\color{rgb,1:red,0.6;green,0.6;blue,0.6}172} & {\color{rgb,1:red,0.35;green,0.35;blue,0.35}157} & {\color{rgb,1:red,0.6;green,0.6;blue,0.6}172} & {\color{rgb,1:red,0.35;green,0.35;blue,0.35}156} & {\color{rgb,1:red,1.0;green,0.5583632666810756;blue,0.19818011428336807}152} & {\color{rgb,1:red,0.7127015300411498;green,0.204225299979703;blue,0.0}137} & {\color{rgb,1:red,0.3277250527283637;green,0.7533866722110684;blue,1.0}150} & {\color{rgb,1:red,0.0;green,0.41394354779309744;blue,0.7090815962463488}\textbf{135}} \\
\arrayrulecolor{lightgray}\midrule\arrayrulecolor{black}
1.55 & {\color{rgb,1:red,0.6;green,0.6;blue,0.6}174} & {\color{rgb,1:red,0.35;green,0.35;blue,0.35}153} & {\color{rgb,1:red,0.6;green,0.6;blue,0.6}154} & {\color{rgb,1:red,0.35;green,0.35;blue,0.35}135} & {\color{rgb,1:red,1.0;green,0.5583632666810756;blue,0.19818011428336807}132} & {\color{rgb,1:red,0.7127015300411498;green,0.204225299979703;blue,0.0}117} & {\color{rgb,1:red,0.3277250527283637;green,0.7533866722110684;blue,1.0}7} & {\color{rgb,1:red,0.0;green,0.41394354779309744;blue,0.7090815962463488}\textbf{6}} \\
\bottomrule
\end{tabular}

\end{figure*}

In this section, we present diagrams showing quality/performance efficiency and
the quality as a function of magnification corresponding to the main paper's Figures~\ref{fig:paretoDLSS} and \ref{fig:quality_zoom},
but without denoising the evaluated sequences. The results are shown in Figures~\ref{fig:pareto}
and~\ref{fig:supp_quality_zoom}. Furthermore, in
Figure~\ref{fig:rotation_max_error} and Table~\ref{tab:rotating_errors},
we examine how the maximum
image error differs between the considered algorithms.

We start by noting that the same methods lie on the Pareto frontier,
independent of whether the sequences are denoised (Figures~\ref{fig:paretoDLSS} and \ref{fig:pareto}).
Furthermore, in Figure~\ref{fig:supp_quality_zoom}, as also shown in Figure~\ref{fig_lookups},
we again see that our algorithms provide perfect bilinear filtering
above a magnification threshold (1.59 for \WM and 2.35 for \WB),
while the \STF~\cite{Pharr2024} and \WC~\cite{Wronski2025}
algorithms are unable to achieve that for any magnification factor due to their inherently stochastic nature.
At lower magnification factors, where our algorithms are unable
to reach perfect filtering and must use fallback alternatives, they still give higher quality than the prior techniques. (A comparison between our fallback methods and the state-of-the-art algorithms is included
in Section~\ref{sec_fallback_evaluation}.)
Furthermore, both Figures~\ref{fig:pareto} and~\ref{fig:supp_quality_zoom} show
that \WM yields higher-quality results than \WB. We also confirm our earlier findings that
the \WM quality increase comes at a performance trade-off, as \WB is the faster of the two.

Finally, we consider the plots in Figure~\ref{fig:rotation_max_error}
and the results in Table~\ref{tab:rotating_errors}, where the plots show
the \emph{maximum} errors produced by the algorithms we consider,
and the table shows the averages of the curves in those plots. We generated these results using
the same simple magnification setup as explained in Section~\ref{sec:scenes},
with the quad rotating between $[0,90]$ degrees for each measurement.
For low magnification, Box and \WM perform the same
(the \WB curves in Figure~\ref{fig:rotation_max_error_115} are offset slightly for visibility).
As magnification increases, \WM starts producing lower errors
than \WB, when both use the same fallback method. At a magnification factor of 1.55,
\WM is able to produce perfect bilinear filtering for most rotations, while \WB requires
higher magnification to do so (see Figure~\ref{fig_lookups}).
The more complex fallback method (\textbf{C+}) often
give lower maximum errors than the simpler, faster one.
These results agree with Figure~\ref{fig:supp_quality_zoom}.

\subsection{Convergence Analysis}
\begin{figure}[t]
	\centering
	\includegraphics[width=\columnwidth]{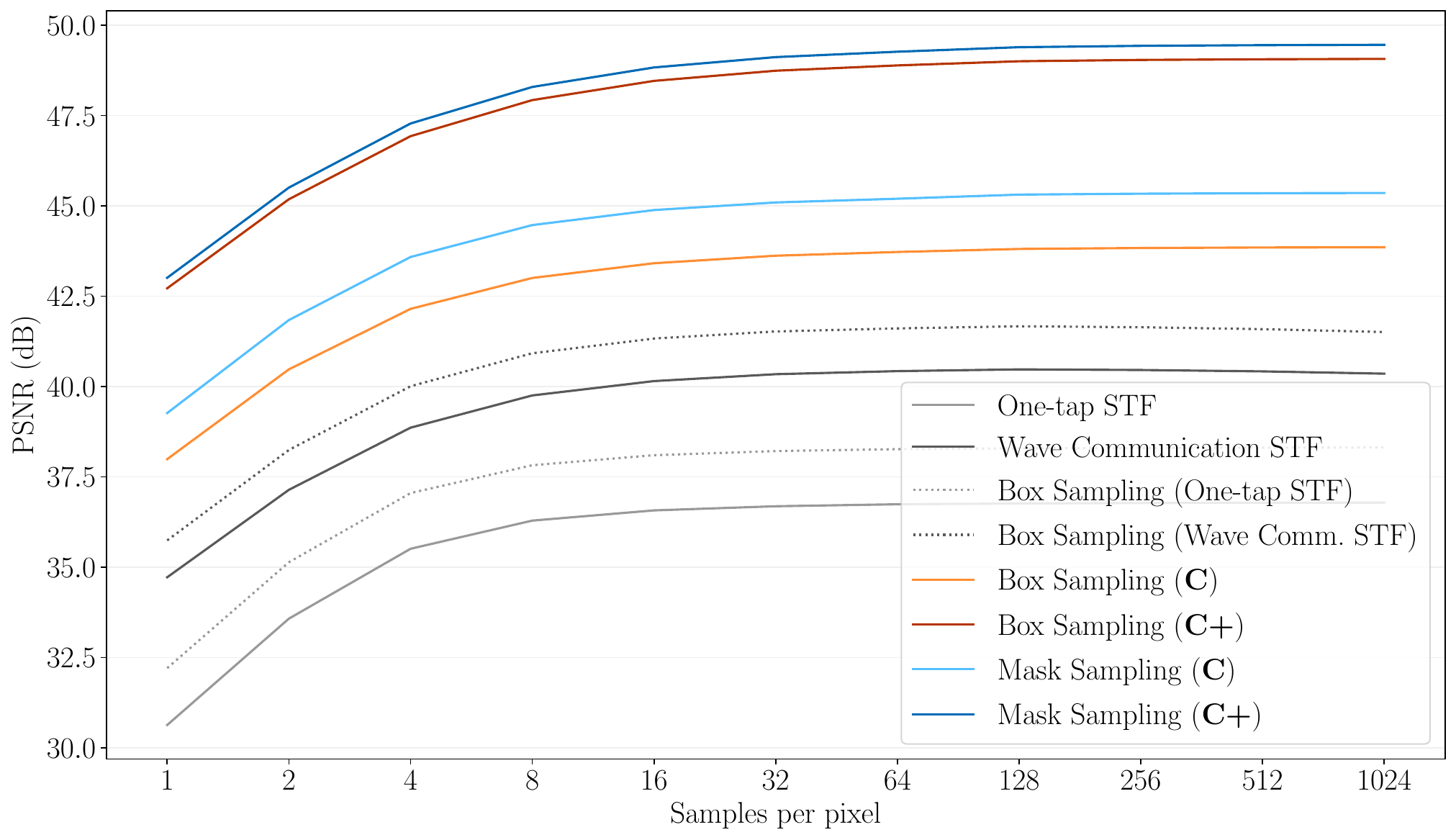}
	\vspace{0.75em}
	\caption{Plot of the error convergence of the considered algorithms when compared to perfect bilinear filtering.
		The scene used was the first scene described in Section~\ref{sec:scenes}.
		Ten magnification factors in the range $[1.0, 2.5]$ were used for each PSNR value.
	}
	\label{fig_spp}
\end{figure}
We also analyzed whether or not the output of our algorithms converges
to the ground-truth image when we increase the number of samples drawn per pixel (SPP),
or how large the bias is if they do not; see Figure~\ref{fig_spp}, which
shows PSNR as a function of SPP.
For these results, we once again use
the same simple, rotated quad and set of textures described in Section~\ref{sec:scenes}.
For a given SPP-value,
PSNR was computed over a set of magnification factors in the $[1.0, 2.5]$ range, as that is approximately the range
where one or more of our algorithms are unable to produce perfect bilinear filtering (see Figure~\ref{fig_lookups}).
Figure~\ref{fig_spp} shows that
our algorithms achieve better results, though they also show some bias,
similar to \STF~\cite{Pharr2024} and \WC~\cite{Wronski2025}.
This is a consequence of the difference between filtering after shading and
filtering before shading, as discussed by Pharr et al.~\cite{Pharr2024}.

\subsection{Fallback Evaluation}
\label{sec_fallback_evaluation}
\begin{figure}[b]
	\centering
	\begin{subfigure}{0.48\textwidth}
		\includegraphics[width=\textwidth]{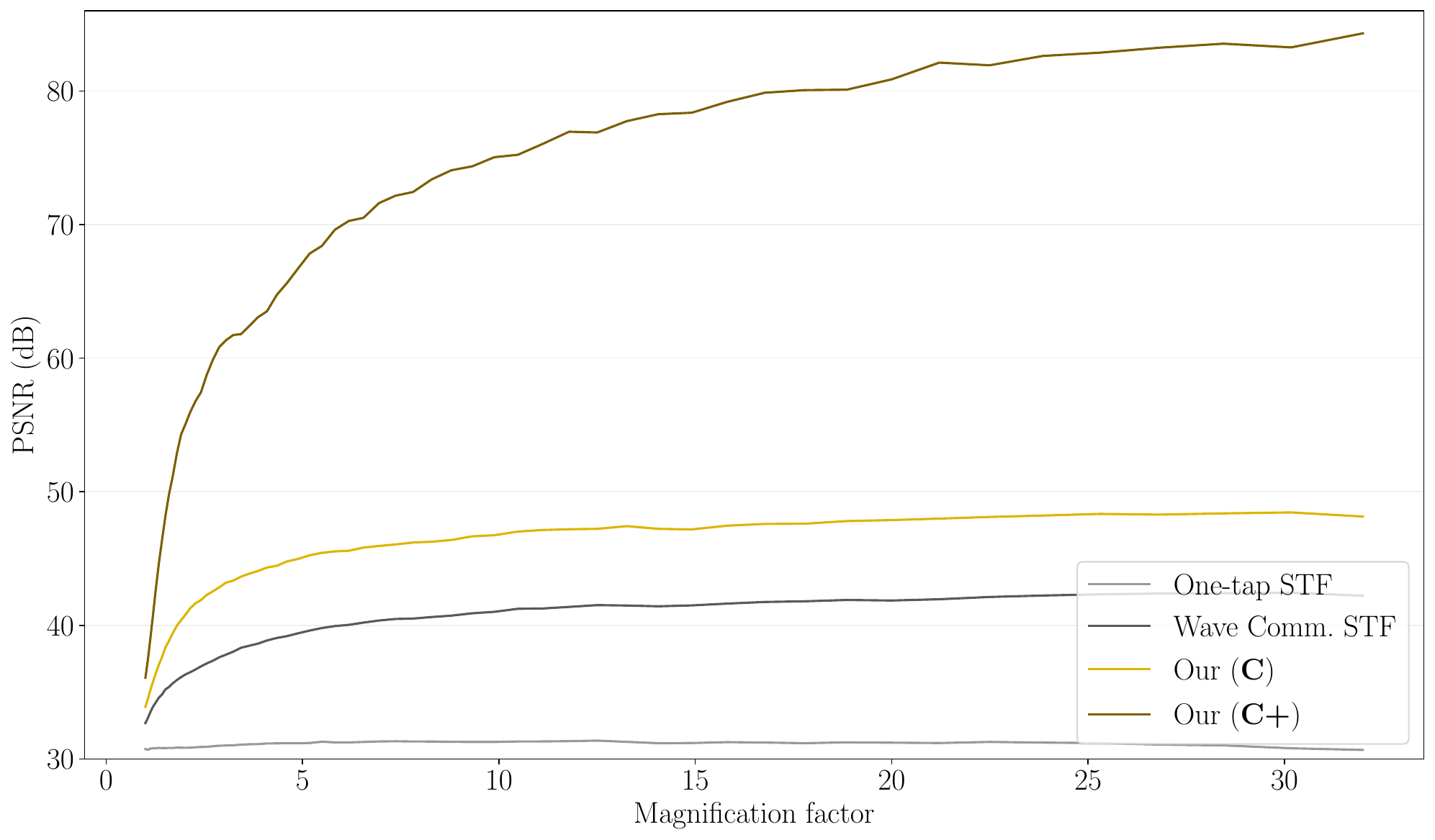}
		\caption{Not denoised.}
		\label{fig_fallback_no_DLSS}
	\end{subfigure}\\
	\begin{subfigure}{0.48\textwidth}
		\includegraphics[width=\textwidth]{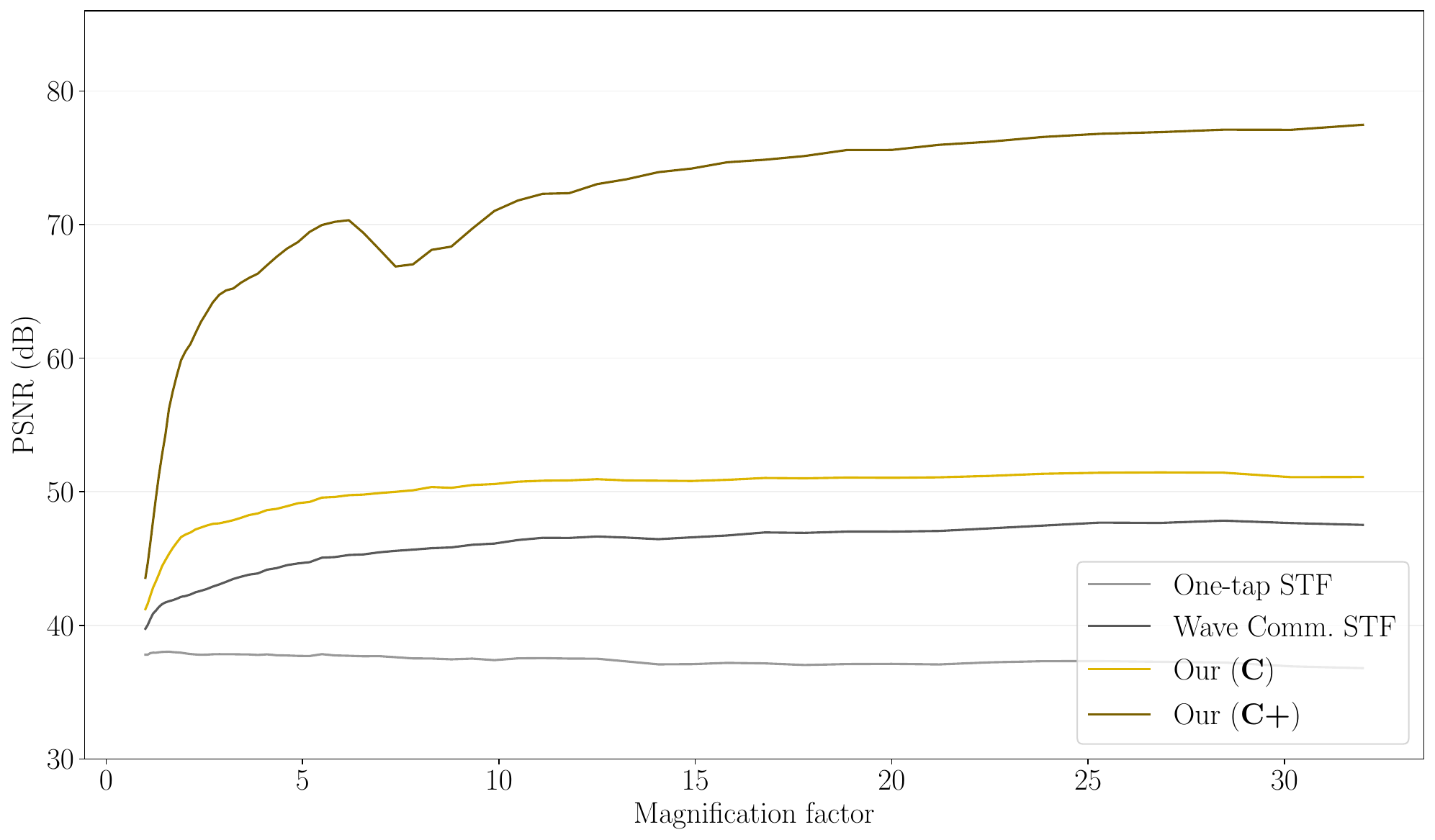}
		\caption{Denoised.}
		\label{fig_fallback_DLSS}
	\end{subfigure}
	\vspace{0.75em}
	\caption{Evaluation of image quality as a function of magnification factor for
		all the different fallback methods. For both without and with denoising (DLSS~\cite{NVIDIA2025:DLSS}),
		our algorithms provide superior quality compared to the other methods.
	}
	\label{fig_fallback_evaluation}
\end{figure}

In extreme cases, every wave 
could contain a silhouette
edge, which means that the fallback (Section~\ref{sec_fallback_method})
would be called for every pixel.
We therefore present
image quality results for our fallback methods
as a function of magnification factor
in Figure~\ref{fig_fallback_evaluation}, both
without and with denoising.

Even our simplest proposed fallback (\textbf{C}) provides better
image quality than both \STF~\cite{Pharr2024} and \WC~\cite{Wronski2025}.
We also note that without DLSS (top diagram),
our \textbf{C+} method continues to increase the image quality
for higher magnification factors, while that happens to a lesser
extent for the other methods. This is a consequence of 
\textbf{C+} leveraging unused lanes to produce more unique texels
and as magnification increases there are more and more
unused lanes that can be used to increase image quality.
For the comparisons with denoised sequences, 
since we apply denoising both to the ground-truth and the test sequence
and the latter almost is identical to the former (see the high PSNR
values for high magnification in Figure~\ref{fig_fallback_no_DLSS}),
the small errors become slightly larger, spatially,
due to the slight blur caused by the denoiser. Presumably, this results
in lower PSNR values for the \textbf{C+} method after denoising
compared to before.
Although the dB difference caused by these differences is large,
the per-pixel errors in the denoised sequences are very small:
at high magnification, the largest errors for the \textbf{C+} method
are 1--2 when scaled to the $[0, 255]$ range.
However, we note that the denoised results for \textbf{C+} show
significantly worse quality than the non-denoised ones
around magnification factor 7--8. While inspecting the data,
we found that the quality decrease originated from the \textsc{Rails}
results, where errors became slightly larger for those magnification factors
despite the image content not changing drastically.
This was due to the denoised \textbf{C+} images being marginally more blurry than the ground truth at those magnification factors.
For the other textures, this effect was not observed.

\subsection{Additional Bicubic Results}
\label{sec_bicubic_zoom}
\begin{figure*}[t]
	\centering
	\begin{subfigure}{0.48\textwidth}
		\includegraphics[width=\textwidth]{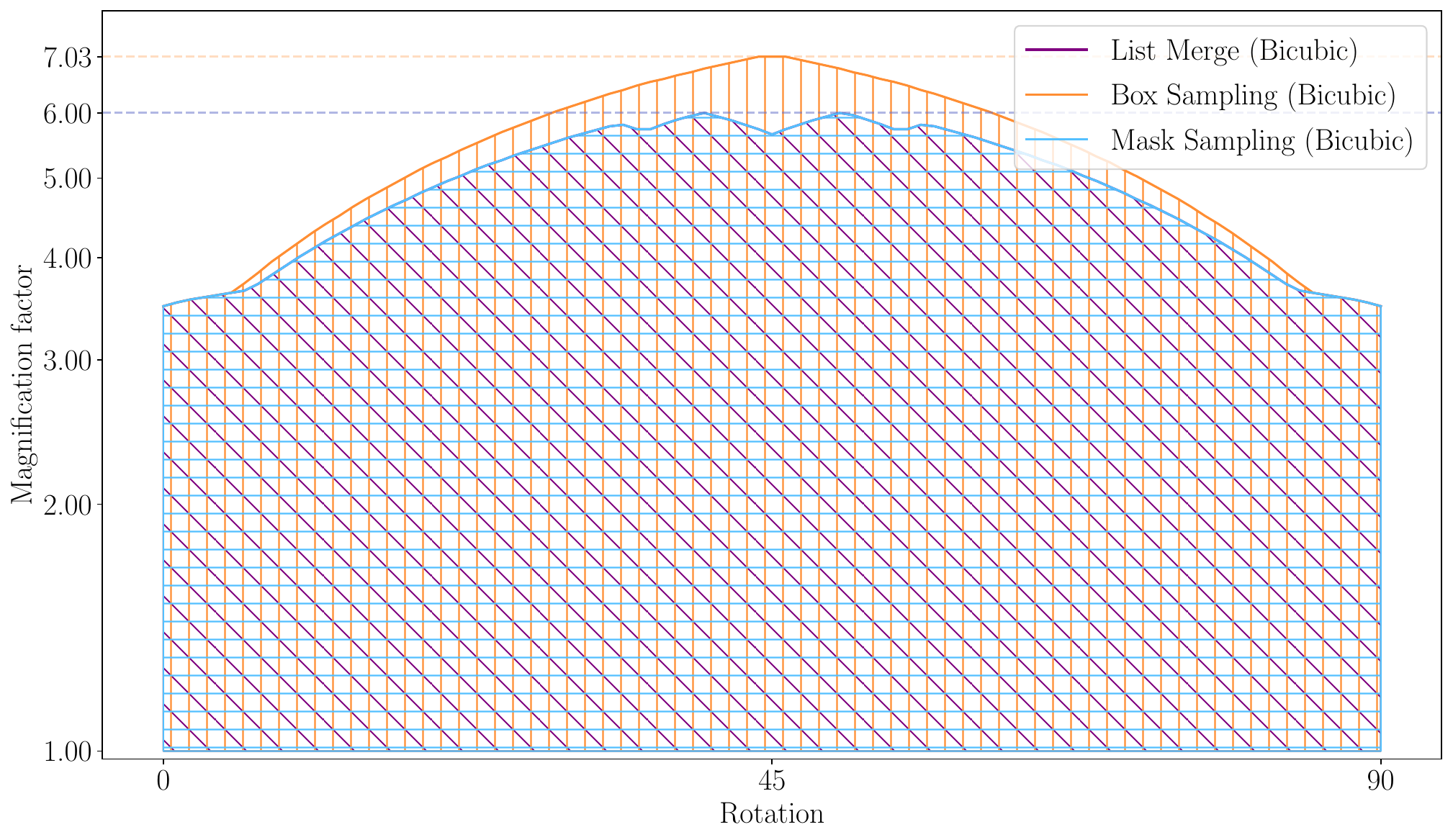}
		\caption{Texel lookups per lane: 1.}
		\label{fig:bicubic_lookups_32}
	\end{subfigure}
	\begin{subfigure}{0.48\textwidth}
		\includegraphics[width=\textwidth]{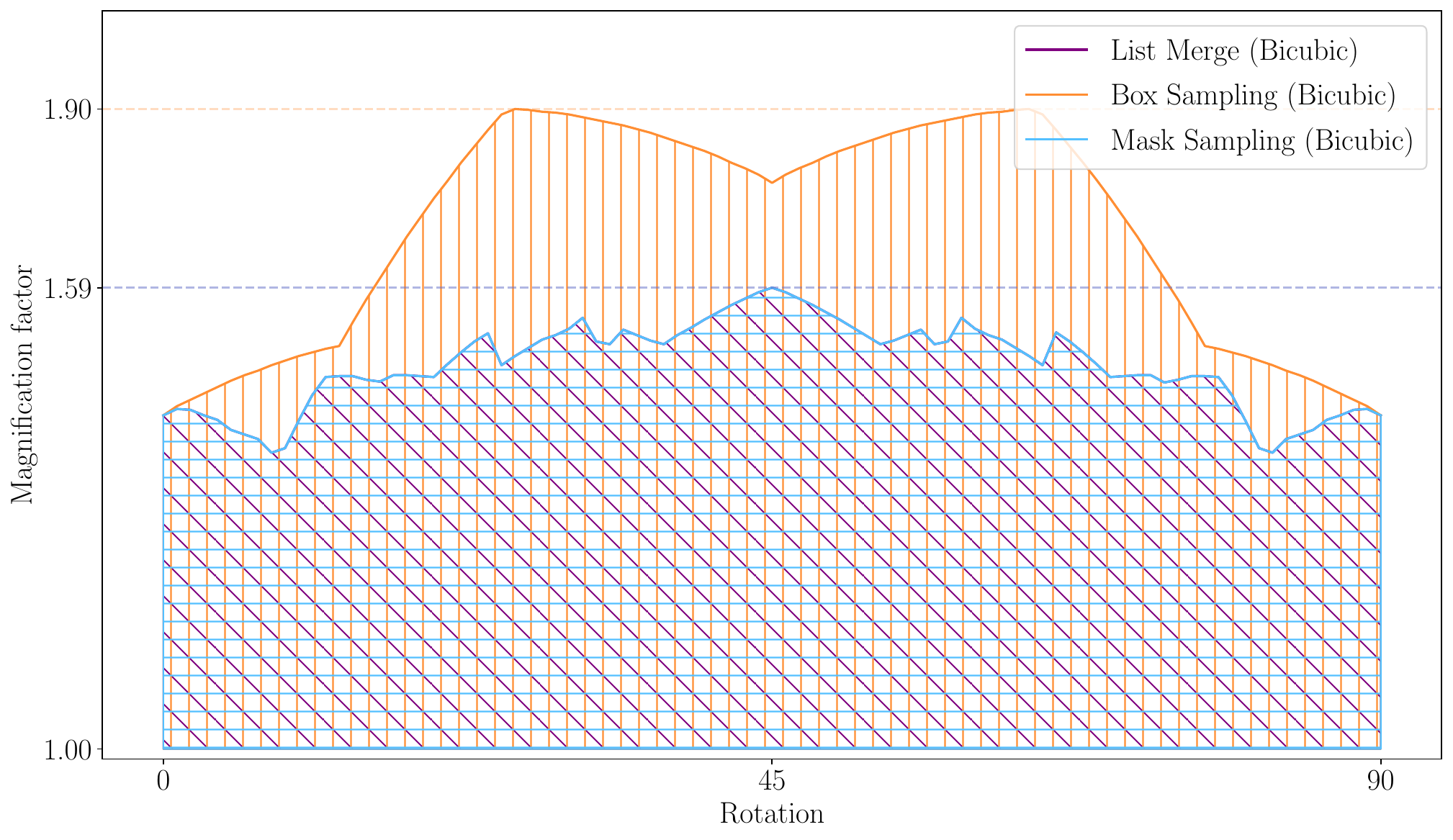}
		\caption{Texel lookups per lane: 2.}
		\label{fig:bicubic_lookups_64}
	\end{subfigure}
	\vspace{0.75em}
	\caption{Illustration of the minimum degree of magnification necessary to achieve perfect
		filtering under different rotations of the quad for perfect bicubic filtering.
		Below those magnification factors, our algorithms need to rely on fallback methods (Section~\ref{sec_fallback_method}).
		Cases where a fallback was necessary are indicated by colored areas.
		The scene used was the first one described in Section~\ref{sec:scenes}.		
		The results in the left diagram are based on producing no more than one texel
		per lane, while those in the right diagram allow up to two texels per lane.
		Notice that \WL and \WM both cover identical areas in both figures. Like
		in the bilinear case, the success rate of \WB is lower than that of the other two
		techniques.
	}
	\label{fig:bicubic_lookups}
\end{figure*}
Finally, we measured the amount of magnification that is needed to achieve perfect
bicubic filtering with each of \WL, \WB, and \WM, given 32 lanes per wave.
Figure~\ref{fig:bicubic_lookups_32} is the bicubic counterpart to
Figure~\ref{fig_lookups}. That is, it shows how much magnification
is needed at different rotations of a quad
to achieve perfect bicubic filtering with 32 or fewer texture lookups per wave.
Figure~\ref{fig:bicubic_lookups_64} is similar, but assumes that we
are able to do two texture lookups per lane, so that we instead need 64 or fewer
lookups per wave. The figures show that perfect bicubic filtering
with our methods requires significantly more magnification
if we only allow one lookup per pixel. If two
lookups are allowed, we see that the required magnification factor is closer
to that for perfect bilinear filtering, despite the four times larger filter area.

\end{document}